\documentclass[structabstract]{aa}  
\usepackage{graphicx}
\usepackage{txfonts}
\usepackage{natbib}
\usepackage{longtable}
\usepackage{lscape}
\usepackage{url}
\bibpunct{(}{)}{;}{a}{}{,} 
\begin{document}
\title{Detailed abundance study of four s-process enriched post-AGB
  stars in the Large Magellanic Cloud. \thanks{based on observations collected with the Very Large Telescope
  at the ESO Paranal Observatory (Chili) of programme number 082.D-0941}}
\titlerunning{The s-process in the LMC}

\author{Els van Aarle\inst{1}, Hans Van Winckel\inst{1}, Kenneth De
  Smedt\inst{1}, Devika Kamath\inst{2}, Peter R. Wood\inst{2}}

\institute{Instituut voor Sterrenkunde, K.U. Leuven, Celestijnenlaan
  200D bus 2401, B-3001 Leuven, Belgium \and Research School of
  Astronomy and Astrophysics, Mount Stromlo Observatory, Weston Creek,
  ACT 2611, Australia}

\offprints{H. Van Winckel, Hans.VanWinckel@ster.kuleuven.be}
\date{}

 
\abstract
{The photospheric abundances of evolved solar-type stars of different
  metallicities serve as probes into stellar evolution theory.}
{Stellar photospheres of post-asymptotic giant branch (post-AGB) stars
  bear witness to the internal chemical enrichment processes,
  integrated over their entire stellar evolution. Here we study
  post-AGB stars in the Large Magellanic Cloud (LMC). With their known
  distances, these rare objects are ideal tracers of AGB
  nucleosynthesis and dredge-up phenomena.}
{ We used the UVES spectrograph mounted on the Very Large Telescope
  (VLT) at the European Southern Observatory (ESO), to obtain
  high-resolution spectra with high signal-to-noise (S/N) of a sample
  of four post-AGB stars. The objects display a spectral energy
  distribution (SED) that indicates the presence of circumstellar
  dust. We perform a detailed abundance analysis on the basis of these
  spectra.}
{All objects are $\mathrm{C}$-rich, and strongly enhanced in s-process
  elements. We deduced abundances of heavy s-process elements for all
  stars in the sample, and even found an indication of the presence of
  $\mathrm{Hg}$ in the spectrum of one object. The metallicity of all
  stars except J053253.51-695915.1 is considerably lower than the
  average value that is observed for the LMC. The derived luminosities
  show that we witness the late evolution of low-mass stars with
  initial masses close to 1~M$_{\odot}$. An exception is
  J053253.51-695915.1 and we argue that this object is likely a
  binary.}
{We confirmed the correlation between the efficiency of the
  third-dredge up and the neutron exposure that is detected in
  Galactic post-AGB stars. The non-existence of a correlation between
  metallicity and neutron irradiation is also confirmed and
  expanded to smaller metallicities.  We confirm the status of
  21~$\mu$m stars as post-Carbon stars. Current theoretical AGB models
  overestimate the observed $\mathrm{C}$/$\mathrm{O}$ ratios and fail
  to reproduce the variety of s-process abundance patterns that is
  observed in otherwise very similar objects. Similar results have
  recently been found for a post-AGB star in the Small Magellanic
  Cloud (SMC).}


\maketitle

\section{Introduction}

Post-AGB stars are stars of low and intermediate initial mass ($M \la 7)\,M_{\odot}$)
that
approach the end of their stellar evolution. The post-AGB evolutionary phase
starts when the strong dusty mass
loss, that is thought to characterise the end of the AGB, has ceased. 
The central star will evolve rapidly to the blue of the HR diagram. Here we concentrate on the
photospheric chemical abundances
and want to use the photosphere of post-AGB stars as
ideal tracers to study AGB nucleosynthetic and dredge-up processes. 

Among the known Galactic post-AGB stars, the variety in observed
chemical patterns in the photospheres is large
\citep[e.g.][and references therein]{vanwinckel03}. Some objects are
strongly enhanced in neutron capture (s-process) elements, while others are not enhanced
at all and might even display s-process deficiencies. This bifurcation
is rather strict, as only a limited number of post-AGB objects are
known to be mildly enriched in s-process elements \citep[e.g.][and
references therein]{sumangala11}. This large chemical variety and the
rather small sample of known Galactic post-AGB stars, makes it hard to
understand how these results can be interpret in the framework of
chemical AGB evolution models.

One of the problems with the current sample of known Galactic post-AGB
stars is that their distances, and hence luminosities and initial masses,
are badly determined as practically none of these objects has a reliable
parallax measurement. We want to overcome this problem by studying post-AGB
objects in other galaxies, and therefore focus in this contribution on
the LMC. At a distance of about 50~kpc \citep{storm11} and at a
favourable aspect angle of 35$^{\circ}$ \citep{vandermarel01}, the LMC is far
enough away that all objects can be assumed to reside at the same distance,
and still close enough to allow the observation of individual objects
in detail.

Observations of post-AGB stars can help us interpret the AGB phase of which they are the
progeny. Studying optically bright post-AGB stars holds several
advantages over studying the AGB stars themselves. First, a
simultaneous study of the photosphere and the circumstellar
environment is possible when the circumstellar envelope becomes
optically thin by expansion. Photospheric spectra of AGB stars are veiled by molecular
lines which impedes spectral analysis of trace
elements \citep[e.g.][]{abia08}. Spectra of post-AGB stars are dominated by atomic
transitions rather than molecular transitions and this allows abundance determinations of a
wide range of elements. Furthermore, the photospheres of post-AGB
stars are convenient to study as the spectral types tend to cluster 
in a range where model atmosphere analysis in
the optical is the most reliable.

We initiated a large project to study quantitatively our newly
identified, large sample of post-AGB stars in the LMC
\citep{vanaarle11} and Small Magellanic Cloud (SMC)
\citep{kamath11}. In this paper, we will present the abundance
analysis of four LMC post-AGB stars that were selected from the
70~spectroscopically confirmed post-AGB stars in the catalogue of
optically bright post-AGB stars of the LMC
\citep[see][]{vanaarle11}. We introduce the objects in
Sect.~\ref{sec:samplestars}. The high-resolution UVES spectra and
their reduction is discussed in Sect.~\ref{sec:observations}. 
We performed a detailed abundance analysis of the
obtained spectra which is presented in Sect.~\ref{sec:spectralanalyses}. The
results of this analysis are discussed in the next sections: In
Sect.~\ref{sec:lum_initmass} we use the acquired atmosphere models to
constrain the luminosities and initial masses of our program stars,
the neutron exposure of the present sample is compared to that of some
Galactic post-AGB objects in Sect.~\ref{sec:neutrexp},
Sect.~\ref{sec:J053253extr} deals with the possibility that one of the
objects in the sample is a binary. A first
comparison with chemical AGB models is presented in
Sect.~\ref{sec:agbmod}. We end with some conclusions and future
prospects in Sect.~\ref{sec:concl}.

\section{The sample stars}\label{sec:samplestars}

Basic information on four objects is gathered in
Table~\ref{table:samplestars}. This table lists the full name of each
object, which we will abbreviate in the
rest of this paper to increase the readability.

\begin{table}
\caption{Names and spectral types of the programme stars \citep{vanaarle11}.}  
\centering
\label{table:samplestars}      
\hspace{-0.35cm}                     
\begin{tabular}{ll}        
\hline\hline                 
Object Name & Spectral Type\tablefootmark{a}  \\
(IRAC)  &  \\
\hline                        
J050632.10-714229.8 &  A3Iab/F0Ibp(e) \\
J052043.86-692341.0 &  F5Ib(e)	 \\
J053250.69-713925.8 &  F6Ia	         \\
J053253.51-695915.1 &  G5		 \\
\hline                                   
\end{tabular}
\tablefoot{\tablefoottext{a}{The '/' indicates that more than one low resolution, optical spectrum was obtained.}}
\end{table}

Three objects have a clearly double peaked SED
(Fig.~\ref{fig:SEDnew}): in which the peak at longer wavelengths is
indicative of a freely expanding, detached shell
\citep{vanaarle11}.  In the SED of J053253, the fourth object, only a small,
relatively hot infrared excess is visible.

All objects in our sample display photometric variability. J052043 and
J053250 are semi-regular variables (SRVs) with an unusual warm
spectral type, while J050632 and J053253 are population II Cepheids
with a very low amplitude and periods of 49.1 and 91.04 days
respectively \citep{vanaarle11}. All objects have a peak-to-peak
photometric amplitude of less than 0.1~mag.

Only one object in our sample has been studied before. \cite{volk11}
obtained Spitzer Space Telescope mid-infrared spectroscopy for J052043
and detected a strong 21~$\mu$m feature. The carrier of this dust
feature is still unknown but it is only detected in post-AGB stars
\citep[e.g.][]{hrivnak09}. For the Galactic sample of post-AGB stars,
it is known that the presence of this feature accompanies a central
star which is enriched in carbon and shows a strong enhancement in
s-process elements \citep{vanwinckel00, reddy02, reyniers03,
  reyniers04}. In this paper, we will explore if this remains true for
this object in the LMC.

\section{Observations}\label{sec:observations}

\begin{table}
\caption{Observational log.}             
\label{table:obslog}      
\centering                          
\begin{tabular}{clcccc}        
\hline\hline                 
Object  & Date & Exp. Time & \multicolumn{3}{c}{S/N} \\
 &  & (s) & Blue & Red$_\mathrm{low}$ & Red$_\mathrm{up}$ \\
\hline                        
J050632 & 4-5/11/2008   & 2 x 3005 & 50 & 105 & 110 \\
J052043 & 21-22/11/2008 & 1 x 3005 & 10 &  50 &  60 \\
	& 4-5/12/2008\tablefootmark{a}    & 2 x 3005 & 15 &     &     \\
	&               & 3 x 3005 &    &  80 & 100 \\
J053250 & 17-18/11/2008 & 3 x 3005 & 15 &  90 & 120 \\
J053253 & 19-20/10/2008 & 4 x 3005 & 10 &  50 &  70 \\
\hline                                   
\end{tabular}
\tablefoot{We list the names, dates, exposure times and
  signal-to-noise per pixel of all UVES spectra taken for our sample stars. 'Blue' corresponds to the spectral range from 3280 - 4519~\AA{}, 'Red$_\mathrm{low}$' to 4777 - 5757~\AA{}, and 'Red$_\mathrm{up}$' to 5834 - 6810~\AA{}. 
\\
\tablefoottext{a}{The third blue spectrum had to be removed due to an instrumental error.}}
\end{table}

We obtained high-resolution, high-signal-to-noise UVES spectra
\citep{dekker00} for all stars in our sample.  The
spectra were taken
in service mode in the fall of 2008. The spectrograph splits the light beam from the
telescope into two spectral arms which are operated simultaneously. This
results in a spectral coverage from 3280 to 4519~\AA{} in the Blue
arm, and from 4777 to 5757~\AA{}, and 5834 to 6810~\AA{} in the Red
arm. We used a slit width of
0.8''  which translates into  $\delta\lambda \sim \lambda/$52\,000 in the
blue and $\delta\lambda \sim \lambda/$48\,000 in the red spectral 
region\footnote{\url{http://www.eso.org/sci/facilities/paranal/instruments/uves/inst/}}.
Some details of the observations can be found in
Table~\ref{table:obslog}. 


The data reduction was performed with the UVES pipeline (version
4.3.0), which included bias correction, flat-fielding, and background,
sky, and cosmic hit correction. Even though the image slicer was used
to obtain our spectra, we used optimal instead of average extraction
to convert the frames from pixel-pixel to pixel-order space. This
produced spectra of similar quality but took better care of the
remaining cosmic hits.

\begin{figure*}
\begin{center}
\resizebox{\hsize}{!}{ \includegraphics{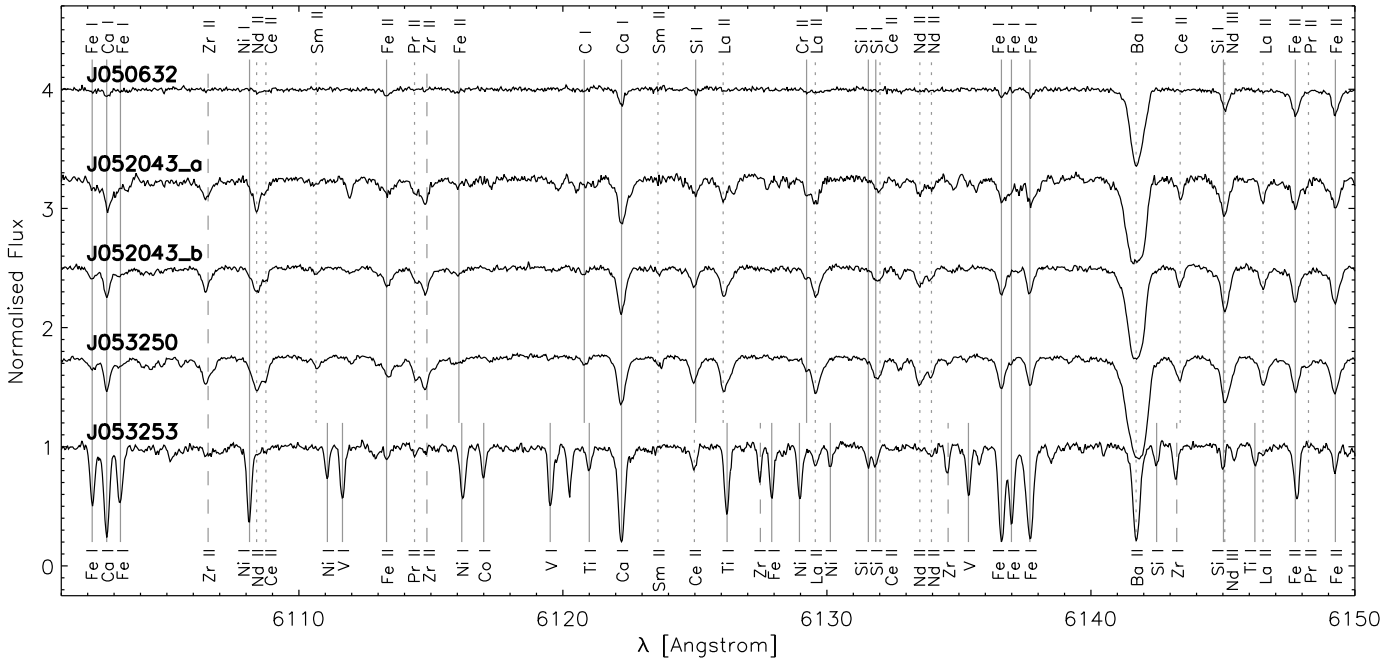}}
\caption[Example of the quality of the spectra of the stars in our sample.]{Example of the quality of the spectra of the stars in our sample. A complete line identification of this spectral interval has been performed using the VALD database \citep{vald}. Elements with atomic number $Z$ smaller than 30 are indicated with a full line, light s-process elements with a dashed line, and heavy s-process elements with a dotted line. As the lower spectrum differs significantly from the upper four, other features are indicated on it. All objects are clearly enhanced in s-process elements, although this is less obvious in the spectrum of J050632 due to its higher temperature.}
\label{fig:spectra}
\end{center}
\end{figure*}

Before merging the different spectra, the barycentric correction of the
spectra, as well as an intrinsic variability check on the different exposures of each
object were performed. All spectra are consistent for all objects,
except for J052043. Its spectra obtained in December show a
small difference in spectral lines with respect to the ones taken in
November. We decided not to merge these spectra,
but to treat them separately. We will indicate the spectrum taken in
November with $a$, and that from December with $b$. We
removed some additional cosmic hits by hand, and merged all consistent
spectra, thus retaining two spectra for J052043 and one for all other
objects. All spectra were normalised by dividing the spectrum by a
spline function defined through interactively identified continuum
points. The reciprocal of the standard deviation
computed on a few normalised continuum windows was used to express the
S/N of our spectra (Table~\ref{table:obslog}) and some sample spectra
are shown in Fig.~\ref{fig:spectra}.

\begin{table}
\caption{Heliocentric radial velocities.}             
\label{table:vrad}      
\centering                          
\begin{tabular}{lcc}        
\hline\hline                 
Object     & $v_r$  & N$_{\mathrm{lines}}$ \\
           & (km/s) &                      \\
\hline                        
J050632    & 218.2 $\pm$ 0.8 & 364 \\
J052043\_a & 243.3 $\pm$ 1.1 & 346 \\
J052043\_b & 244.6 $\pm$ 1.0 & 438 \\
J053250    & 283.3 $\pm$ 0.7 & 278 \\
J053253    & 237.5 $\pm$ 0.8 & 454 \\
\hline                                   
\end{tabular}
\tablefoot{We list the names, radial velocities and the number of lines used to determine these velocities. The errors on the radial velocity are the standard deviations of the individual results of all lines.}
\end{table}

The high resolution and the large wavelength coverage of our UVES
spectra permit us to determine the radial velocities using the
position of a large number of lines.  The line positions were
  measured using a Gaussian fit through the measured profiles. The
  identification of the lines and their rest wavelength were taken 
from the VALD \citep{vald} database. The line identification itself
was performed by comparing the expected line strength of all input
lines using the model atmosphere and overabundances (Sect. 4) of our
chemical analyses. We took automatically the nearest strongest
line. The internal consistancy on a large number of lines 
shows that this method is reliable. The results are listed in
Table~\ref{table:vrad}. The radial velocities we find are consistent
with the average radial velocity of the LMC, which is about 262~km/s
\citep{vandermarel02}.

\section{Spectral Analyses}\label{sec:spectralanalyses}

We use the list of lines that are useful for the chemical analysis of
A, F, and G stars that has been gathered at the Instituut voor
Sterrenkunde and is regularly updated. The initial list is described
in \cite{vanwinckel00}, and contains lines for a range of elements
from $\mathrm{He}$ ($Z=2$) up to $\mathrm{Eu}$ ($Z=63$). With this
list, we ensure that the different chemical analyses are performed
homogeneously and we restrict ourselves to lines with reliable
oscillator strengths. We determined the equivalent
widths of the lines by fitting Gaussian profiles to the observations
to take care of blending in their wings.

To determine the atmosphere parameters and obtain the abundances of
trace elements, we used the ATLAS~9
model atmospheres \citep{castelli04} in combination with the latest
version (August 2010) of the LTE abundance calculation routine MOOG
\citep{sneden73}. The model gridsteps are $\Delta T_{\rm eff} = +/- 250 K$,
$\Delta \log g = +/- 0.5$ and $\Delta$([Fe/H]) = +/- 0.5

\subsection{Atmospheric parameters}\label{ssec:atmod}

We used several spectroscopic tracers to quantify the effective temperature of our
low gravity objects.  A useful upper limit for the effective
temperature is provided by the absence of $\mathrm{He}$ lines in the
spectra. At the S/N of our spectra, this means that the lines must be
weaker than $\sim$ 5 m\AA.  The absence of He lines, is especially
constraining for the hottest object in our sample J050632. 
In Table~\ref{table:he}, we list for several models the expected equivalent width of the 
strongest He features when assuming a solar helium abundance.
The solar He abundance is a lower limit for the
actual abundance as an increase of the helium abundance
in the envelope of our stars of up to 20~\% can be expected during the
AGB evolution \citep[e.g.,][]{cristallo09, cristallo11}.

\begin{table}
\caption{Expected equivalent widths of the strongest $\mathrm{He}$ lines (at 4471.50, 5875.65 and 5876.50~\AA{}), and iron abundances for the best atmosphere model for a range of effective temperatures for J050632.}             
\label{table:he}      
\centering                          
\begin{tabular}{cccccc}        
\hline\hline                 
$T_{\mathrm{eff}}$ & \multicolumn{3}{c}{Expected EW (m\AA{})} & $\log \epsilon_{\mathrm{Fe I}}$\tablefootmark{a} & $\log \epsilon_{\mathrm{Fe II}}$\tablefootmark{a}\\
         (K)       & 4471.50 & 5875.65 & 5876.50 & (dex)           & (dex) \\
\hline                        
6250               &   0.7         &   0.3         &   0.3         & $5.69 \pm 0.23$ & $6.20 \pm 0.23$ \\
6500               &   4.6         &   1.1         &   1.2         & $5.89 \pm 0.20$ & $6.25 \pm 0.20$ \\
6750               &  10.5         &   2.8         &   3.0         & $6.09 \pm 0.18$ & $6.28 \pm 0.15$ \\
7000               &  17.6         &   5.6         &   6.0         & $6.34 \pm 0.17$ & $6.37 \pm 0.12$ \\
\hline                                   
\end{tabular}
\tablefoot{\tablefoottext{a}{The listed errors are the line-to-line
    scatter. The \ion{Fe}{I} abundance is based on the measurements of 63 lines and that of \ion{Fe}{II} on 30.}}
\end{table}

An excitation analyses of Fe I is a traditional temperature sensor.
Towards lower metallicities, higher effective temperatures, and lower
gravities, the increasing UV radiation field and decreased rate of
electron collisions may create an over-ionisation and over-excitation of iron
with respect to the LTE description \citep{rentzschholm96, mashonkina11}. 
The same effect can be
found in the abundances of other neutral elements, especially those of
the iron peak. Observationally,
these non-LTE effects have been noted in, amongst others, the
classical Cepheid $\delta$~Cep \citep{kovtyukh99}, and some post-AGB
stars \citep[][]{takeda07, gorlova12}. This means that the Teff we estimate from the excitation analyses 
of Fe I may be too high. At these temperatures this photoionisation scarcely
alters the abundance of \ion{Fe}{II}, with differences of less than
0.01~dex \citep{mashonkina11} in the relevant temperature regime of
our programme stars. The effective temperature deduced from a general
excitation analysis of the \ion{Fe}{II} lines is, however, less
accurate because of the smaller number of available features and the
lower spread in excitation potentials in optical \ion{Fe}{II} lines.

Hydrogen line profiles do not show a strong dependence on
metallicity of the stellar photosphere and we obtained the effective
temperature of our stars by comparing their Balmer profiles with the
synthetic spectra of \cite{coelho05}.
When estimating the appropriate effective temperature, we focused on the
shape of the wings, as the depth of the feature may deviate from the
synthetic spectra. This is the case because the core of the line is
formed in the surface layers of the photosphere where the standard
atmosphere models are less accurate due to the occurrence of surface
phenomena and the boundary conditions of the models. We assume
that the temperature and density gradient of our objects are similar
to a model in hydrostatic equilibrium, despite the fact that the stars are pulsating.

For the given effective temperature, we first
determined $\log g$ by imposing ionisation equilibrium for iron and,
when available, other elements with lines of diferent ions. Only
lines with an equivalent width of less than 150~m\AA{} were used, as
the equivalent width of stronger lines is no longer strongly
correlated with the abundance of the element, because they lay on the
flat part of the curve of growth. We tried to estimate the \ion{Fe}{I}
abundance by extrapolating it to an equivalent width of 0~m\AA{} as
described in \cite{kovtyukh99}, but the relatively large scatter on
abundance and equivalent width of our lines rendered this
impossible. The average abundance was used instead. We changed the
metallicity of the model according to the obtained \ion{Fe}{II}
abundance. Finally, we estimated the microturbulent velocity $\xi_t$ in
steps of 0.5~km/s by demanding that the abundance derived for the
individual \ion{Fe}{II} features does not depend on their reduced
equivalent width. This procedure was repeated until the results
converged towards a single model.

\begin{figure}
\begin{center}
\resizebox{\hsize}{!}{ \includegraphics{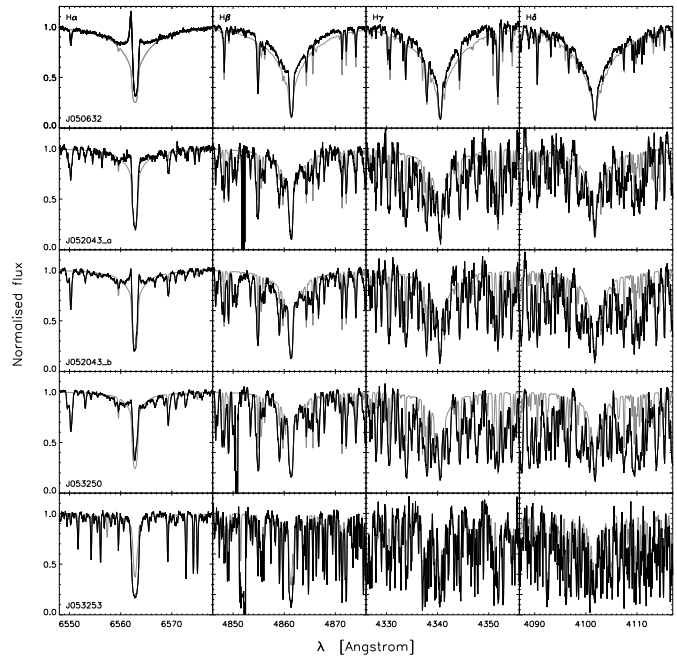}}
\caption[Comparison with the synthetic spectra of \cite{coelho05}.]{Comparison with the synthetic spectra of \cite{coelho05}. We show for each spectrum the first four Balmer lines and the synthetic spectrum of the final atmosphere model (see also Table~\ref{table:atmod}). All synthetic spectra have been convolved with a Gaussian to match the resolution of the observed.}
\label{fig:balmerall}
\end{center}
\end{figure}

The final atmosphere models for our sample can be found in
Table~\ref{table:atmod} and their corresponding synthetic Balmer
profiles in Fig~\ref{fig:balmerall}. The final atmosphere models for
both spectra of J052043 show a slight discrepancy in preferred
gravities. Because of the coupling of this parameter with the
effective temperature, this indicates a small temperature
difference as well. As models with ionisation equilibrium between \ion{Fe}{I} and
\ion{Fe}{II} require higher values of $\log g$ at higher effective
temperatures, this confirms that J052043 became slightly cooler in
December than it was in November. 
The errors on the different atmosphere parameters are estimated to be
the grid steps and therefore no interpolation of models and/or new
models were computed.

\begin{table}
\caption{Final atmosphere models.}             
\label{table:atmod}      
\centering                          
\begin{tabular}{lcccc}        
\hline\hline                 
Object     & $T_{\mathrm{eff}}$ & $\log g$ & $\xi_t $ & [$\mathrm{Fe}$/$\mathrm{H}$] \\
           & (K)       & (dex)    & (km/s)    & (dex) \\
\hline              
J050632    & 6750      &  0.5     &  3.0      & -1.0 \\ 
J052043\_a & 5750      &  0.5     &  3.0      & -1.0 \\ 
J052043\_b & 5750      &  0.0     &  3.0      & -1.0 \\ 
J053250    & 5500      &  0.0     &  3.0      & -1.0 \\ 
J053253    & 4750      &  2.5     &  2.5      & -0.5 \\ 
\hline                                   
\end{tabular}
\end{table}

\subsection{Spectrum synthesis}

The abundances of some elements could not be determined on the basis
of isolated lines. An estimate of these
abundances was nevertheless obtained through spectrum synthesis. To
match the synthetic line profiles with the observed ones, we need a
good description of the line-broadening which is the combined effect of instrumental,
macroturbulent, and rotational broadening. Because the latter is
thought to be small in F-G supergiants, and its effect on the spectra
is difficult to distinguish from the macroturbulent broadening, we
neglect it in our calculations. The macroturbulent
broadening as the only unknown broadening parameter in the line
profile fitting procedure.

\subsubsection{The macroturbulent broadening $\xi_m$}\label{sssec:macrbr}

\begin{figure}
\begin{center}
\resizebox{9cm}{!}{ \includegraphics{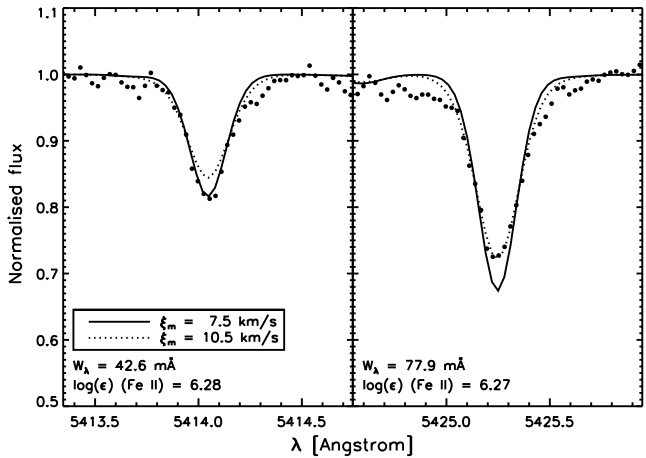}}
\caption[Illustration of the necessity of different values of the macroturbulence at different reduced equivalent widths.]{Illustration of the necessity of different values of the macroturbulence at different reduced equivalent widths. We display two \ion{Fe}{II} lines from the spectrum of J050632 from the same spectral region and with a similar abundance derived from their equivalent width. Their macroturbulent broadening factor differs, however, by 3~km/s.} 
\label{fig:macrov_vb}
\end{center}
\end{figure}

Based on the results \cite{reyniers04} obtained for the
$\mathrm{C}$-rich, Galactic post-AGB star IRAS\,06530-0213, we expect
that a single value for the macroturbulent broadening will not be
sufficient to fit all lines in a spectrum equally well. This is illustrated in
Fig.~\ref{fig:macrov_vb} were we show two \ion{Fe}{II} lines in the
spectrum of J050632 with similar abundances, as derived from their
equivalent width, and from the same spectral region. The two lines
clearly require a different macroturbulent broadening factor. The
reason for this dependence is not clear, but it is suggested that this
is an optical depth effect where the stronger lines are on average
being formed at lower geometrical depth \citep{reyniers04}. As similar
trends have been found in HD\,172481 \citep{reyniers01} and all other
objects in our sample, this behaviour is certainly not
exceptional. Most authors probably do not notice this effect as
usually only a few lines or lines in the proximity of the feature at
hand are investigated to determine the macroturbulent broadening.

We determined the macroturbulent broadening by a $\chi^2$ minimisation
of the difference between a radial-tangential broadening profile
\citep{gray08} that is corrected for the instrumental and microscopic
broadening, and the observed feature.  As the equivalent width
remains unchanged while the line profile is broadened, we used the
derived mean abundance of each element and left the macroturbulent
broadening as the only free parameter to fit the line profile. We
varied the broadening velocity in steps of 0.5~km/s. Only the best,
unblended lines of different elements and ionisation states were used
in this procedure.

\begin{figure}
\begin{center}
\resizebox{9cm}{!}{ \includegraphics{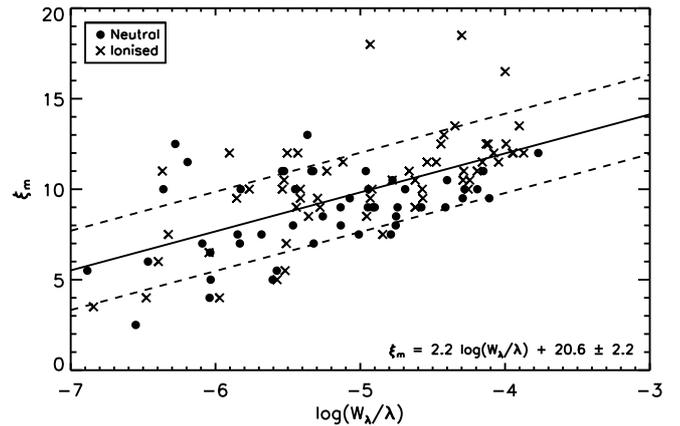}}
\caption[The macroturbulent broadening of J050632 as a function of reduced equivalent width.]{The macroturbulent broadening $\xi_m$ of J050632 as a function of reduced equivalent width $\log(W_{\lambda}/\lambda)$. The dashed lines are the standard deviation on the mean trend.}
\label{fig:macrov_rel}
\end{center}
\end{figure}

\begin{table}
\caption{Fitting parameters of the relation between the macroturbulent broadening and the reduced equivalent width.}             
\label{table:vmacro}      
\centering                          
\begin{tabular}{lccccc}        
\hline\hline                 
Object     & Slope & Intersection & $\sigma$\tablefootmark{a} & Corr. Coeff. & N$_{\mathrm{lines}}$ \\
\hline                        
J050632    & 2.2   &     20.6     &    2.2   &    0.60      & 107 \\
J052043\_a & 5.4   &     30.4     &    2.3   &    0.79      &  21 \\
J052043\_b & 3.5   &     25.1     &    1.9   &    0.78      &  52 \\
J053250    & 3.3   &     22.8     &    1.7   &    0.64      &  30 \\
J053253    & 1.0   &      8.9     &    2.2   &    0.26      &  58 \\
\hline                                   
\end{tabular}
\tablefoot{\tablefoottext{a}{Standard deviation with respect to the fitted trend.}}
\end{table}

The resulting relation between the macroturbulent broadening and the
reduced equivalent width for J050632 is shown as an example in
Fig.~\ref{fig:macrov_rel}. More information on the fitting parameters
of this relation in the other spectra is listed in
Table~\ref{table:vmacro}.  The clear difference for ionised and neutral species that is
observed in \cite{reyniers04}, is not recovered for our spectra and
the slopes we find are much smaller than their slope of 10.9.

%
%
%

\subsubsection{Blending}\label{sssec:specsynth}

All the single lines used to obtain specific abundances were checked individually to
avoid the possibility that unrecognised blends could
cause an overestimate of the abundance.
For all used and presumably single lines, a list of all potential
neighbouring lines was obtained from the VALD \citep{vald} database. Their expected
equivalent widths were computed based on the mean abundances found for
every chemical element. 

For each used blend, a list of all neighbouring lines in an interval of
1~\AA{} was extracted from VALD. We calculated the appropriate
macroturbulent broadening with the relations from
Sect.~\ref{sssec:macrbr} for the total equivalent width of the line,
as the element under consideration is the dominant contributor of the
blend. This induces a possible error, as this relation is calculated
for unblended lines, but it is unclear how it should be adapted in the
case of blended lines. The {\sl synth} driver of MOOG and all known
abundances were used to perform the spectrum synthesis. The final
abundance was obtained by a $\chi^2$ minimisation of the difference
between the observed spectrum and the model in the spectral range used
to calculate the equivalent width of the line.  The abundances of
$\mathrm{S}$,  $\mathrm{Na}$, $\mathrm{Zn}$, $\mathrm{Ba}$,
and $\mathrm{Sm}$ were based on blended lines.

\subsection{Elements beyond the Ba-peak}

\cite{reyniers03b} detected gadolinium, ytterbium, lutetium, and maybe
also tungsten in the spectra of three Galactic, s-process enriched
post-AGB stars.  Inspired by their findings, we performed a specific systematic
search for lines of elements beyond the $\mathrm{Ba}$-peak. We focused on the
elements from gadolinium ($\mathrm{Gd}$, $Z = 64$) up to lead
($\mathrm{Pb}$, $Z = 82$), as the important features of elements with
lower atomic numbers were already included in our original list (see
Sect.~\ref{ssec:atmod}). 

For each element, the atomic data of all
features in the optical spectrum were acquired from VALD
\citep{vald}. The strongest expected lines were determined by imposing an ad
hoc high abundance and calculating the corresponding equivalent widths
for the appropriate atmosphere model. 
We then ordered the lines according to expected relative strength and 
systematically checked the obtained spectra for the presence of the
spectral lines.

We discarding all lines for which (1) the nearest line in the spectrum
was not centred on the given wavelength within 1 km/s, (2) the signal to noise in the spectral
region was too low to recognize a genuine spectral line, or (3) the line was located in a crowded region
where the error on the position of the continuum appeared too
large. For the remaining features, a list of lines of elements with
already known abundance in the interval,
was extracted from VALD and the equivalent width of these features
determined with MOOG. Only lines of which about ninety percent of the
observed equivalent width could not be explained by the elements of
which the abundance was already known at that stage, were kept in the
sample.  After this analyses, we compared the abundances derived from all lines of
the specific element. Features with too high an equivalent width were
discarded as blends with unrecognised contributors. A final quality
check was performed by recalculating the expected equivalent width of
the strongest lines of the element at hand with this new final abundance,
and comparing this to what is seen in the spectra. As the atomic data
of all lines listed in VALD are not equally reliable, we only executed the latter test for
elements of which the abundance is based on at most three lines.



The strongest features of all heavy s-process elements 
are concentrated in the low S/N region from 3280 to
4519~\AA{} and we were plagued by the low S/N ratio of our spectra in
the blue. Nonetheless, we were able to determine the abundances for a range of
heavy s-process elements. Lines of $\mathrm{Gd}(64)$,
$\mathrm{Tb}(65)$, $\mathrm{Dy}(66)$,  $\mathrm{Er}(68)$, $\mathrm{Tm}(69)$, $\mathrm{Yb}(70)$,
$\mathrm{Hf}(72)$, $\mathrm{Ta}(73)$, $\mathrm{W}(74)$,
$\mathrm{Hg}(80)$ were positively itentified in many of the objects,
and the results are given in the abundance Tables~\ref{table:abres1}, \ref{table:abres2},
\ref{table:abres4} and \ref{table:abres5}.

\subsection{Hyperfine structure}


We investigated the influence of hyperfine splitting (hfs = the
splitting of spectral lines caused by nucleon-electron spin
interaction) on the derived abundances of all odd-$Z$ elements from
$\mathrm{La}$ ($Z = 57$) to $\mathrm{Lu}$ ($Z = 71$). Because hfs
causes extra de-saturation, it will only affect the abundance
determination when this is based on strong lines that are situated on
the flat part of the curve of growth.

To make sure our results are independent of the method used for
abundance determination, we first recalculated the abundances of the
elements  by use of spectrum synthesis (see
Sect.~\ref{sssec:specsynth}), treating each feature as a single
line. This process was then repeated taking hfs into account, with the
final goal of comparing the acquired abundances. For each spectral line, we
obtained the energy levels corresponding to its wavelength from the
National Institute of Standards and
Technology\footnote{\url{http://www.nist.gov/pml/index.cfm}}
(NIST). Appropriate values of the hfs constants $A$ and $B$ were
obtained from \cite{lawler01a} for \ion{La}{II} ($Z = 57$), from
\cite{sneden09} for \ion{Pr}{II} ($Z = 59$), from \cite{lawler01b} for
\ion{Eu}{II} ($Z = 63$), from \cite{lawler01c} for \ion{Tb}{II} ($Z =
65$), and from \cite{sneden03} for \ion{Lu}{II} ($Z = 71$). We
  ignored $\mathrm{Pm}$ ($Z = 61$, no stable isotope),  
$\mathrm{Ho}$  ($Z = 67$, no detections) and $\mathrm{Tm}$($Z =
69$, no A,B constants found) in our calculations. Because the electric
quadrupole interaction ($B$ constant) has a much smaller effect on the
line component pattern than the magnetic dipole interaction ($A$
constant), it is often neglected and we set its value to zero if it
could not be found in the literature. The relative strengths of the
different hfs components were calculated with the equations given in
\cite{condon35}.

 A real, but small effect of
at most $\Delta_{\mathrm{ab}} = \log \epsilon_{\mathrm{hfs}} - \log
\epsilon_{\mathrm{no\ hfs}} = -0.10$~dex is noticed in some features,
but falls within the original error margin on the abundances, and is
dwarfed by other uncertainties like e.g., the exact value $\log gf$,
the placement of the continuum or undetected blends.
We conclude that the effect of hfs on the derived abundance is very
small.

\section{Abundance results}

In Tables~\ref{table:abres1}, \ref{table:abres2}, \ref{table:abres4},
and \ref{table:abres5} we present the complete abundance analysis
results of the stars in our sample, the details on all lines used
in this analysis can be found in five catalogues that are available at
the CDS\footnote{via anonymous ftp to cdsarc.u-strasbg.fr
  (130.79.128.5) or via
  http://cdsweb.u-strasbg.fr/cgi-bin/qcat?J/A+A/} and contain the
following information. For each catalogue, Col.~1 gives the name of
the ion a certain line belongs to, Col.~2 contains its rest frame
wavelength, Col.~3 lists its lower excitation potential. The logarithm
of the oscillator strength can be found in Col.~4, Col.~5 gives the
measured equivalent width, and the abundance deduced from this line is
mentioned in Col.~6.

We determined the errors on the [el/$\mathrm{Fe}$] abundances in
Tables~\ref{table:abres1}, \ref{table:abres2}, \ref{table:abres4}, and
\ref{table:abres5} with the method described in \cite{deroo05a}, and
the alternative procedure for estimating the error induced by the
parameter uncertainty of the atmosphere model from \cite{reyniers07c}. To estimate this error, we
calculated the abundance difference for a certain element for our
preferred model and for the most appropriate model to fit our data
with a 250~K higher/lower temperature and 0.5 higher/lower gravity
($\log\,g$). The total uncertainty on the
[el/$\mathrm{Fe}$] abundances can be found in
Tables~\ref{table:abres1}, \ref{table:abres2}, \ref{table:abres4},
and
\ref{table:abres5}. It is the quadratic sum of the uncertainties on the
mean due to line-to-line scatter ($\sigma_{\mathrm{ltl}}$), the
abundance ratio uncertainties induced by the model parameters
($\sigma_{\rm teff}, \sigma_{logg}$), and the uncertainty on the $\mathrm{Fe}$
abundance ($\sigma_{\mathrm{Fe}}$):

\[
\sigma_{\mathrm{tot}} = \sqrt{\left(
    \frac{\sigma_{\mathrm{ltl}}}{\sqrt{N_{\mathrm{el}}}} \right)^2 +
  (\sigma_{teff})^2 + (\sigma_{logg})^2 + \left( \frac{\sigma_{\mathrm{Fe}}}{\sqrt{N_{\mathrm{Fe}}}}\right)^2}. 
\]

\begin{figure}
\begin{center}
\resizebox{\hsize}{!}{ \includegraphics{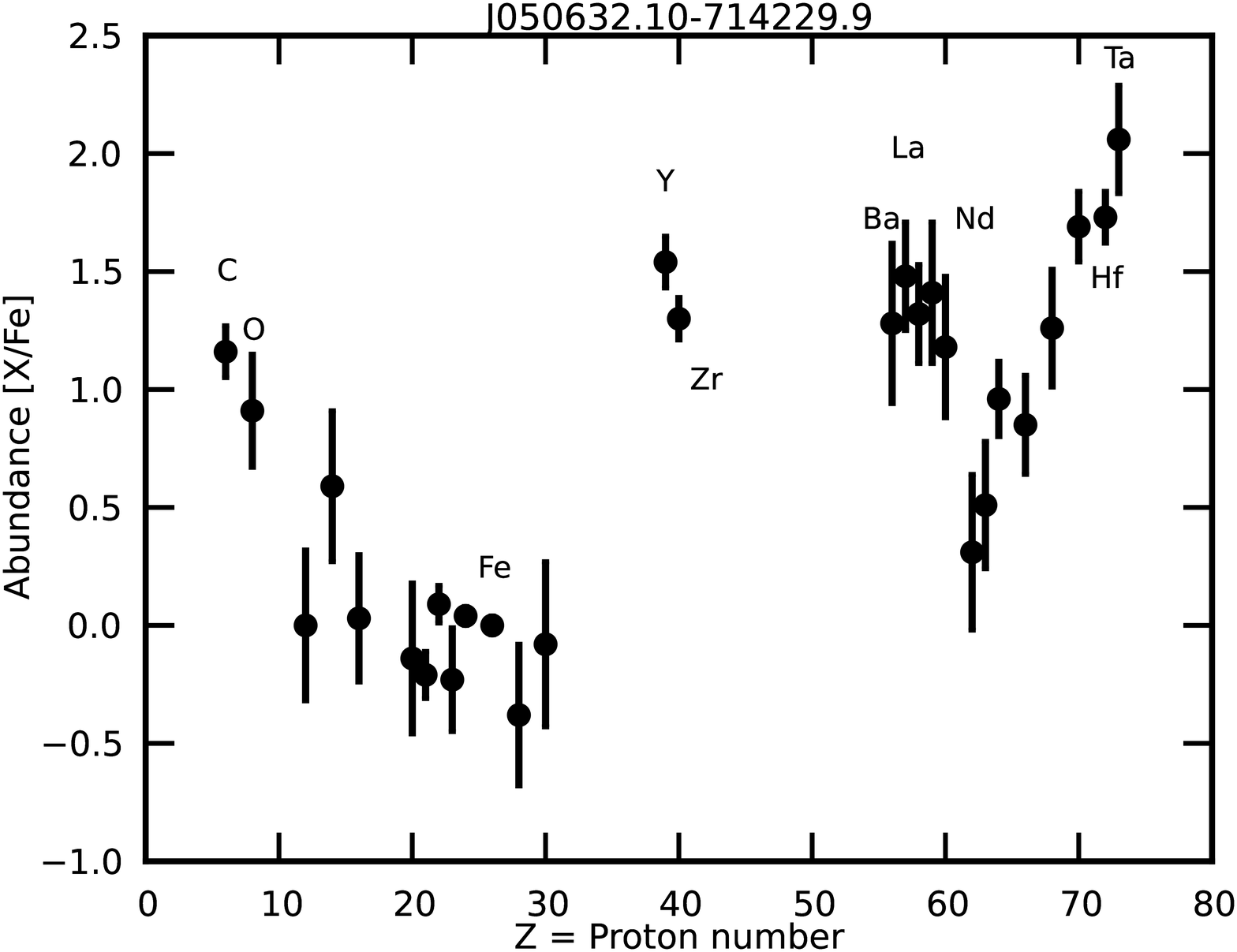}}
\resizebox{\hsize}{!}{ \includegraphics{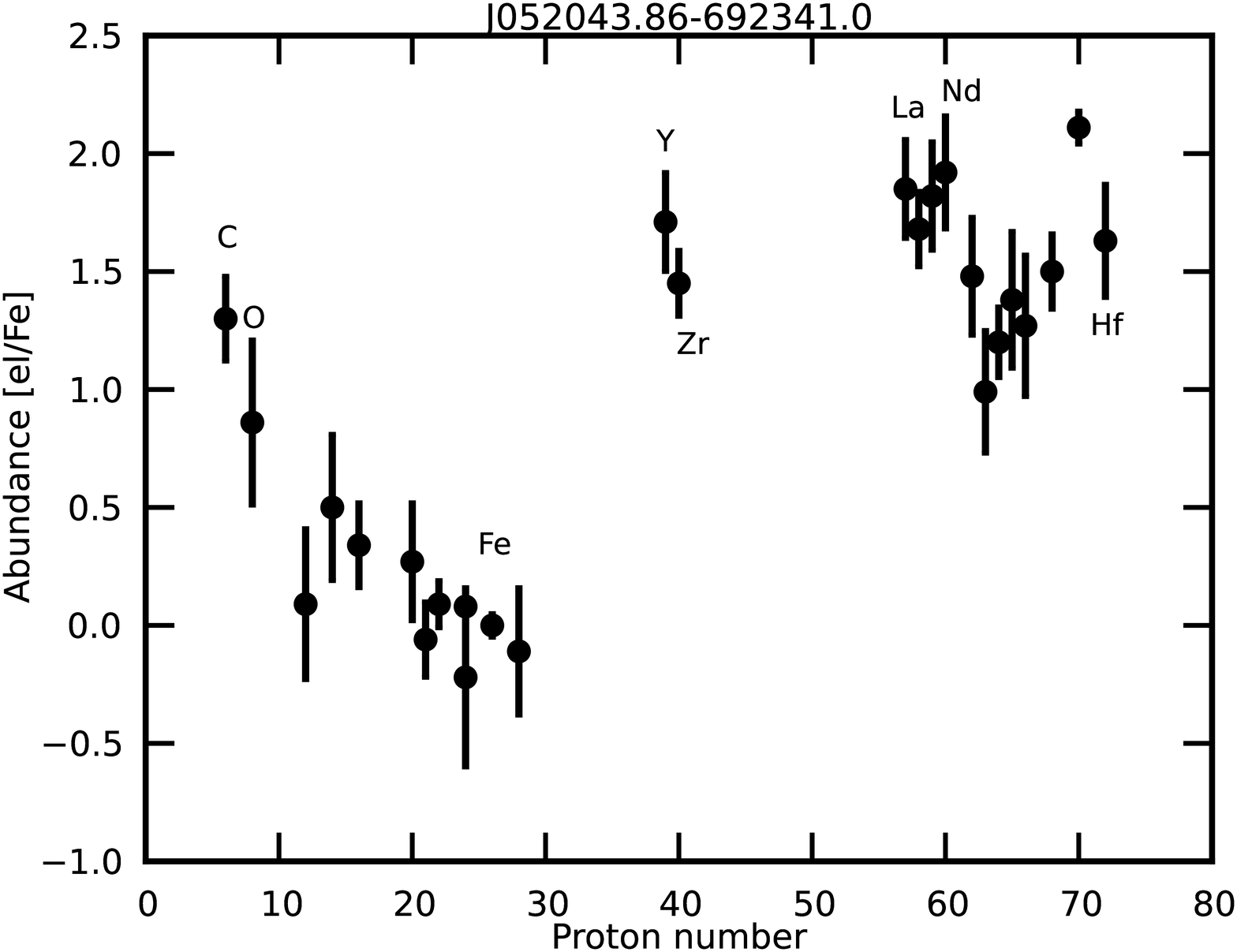}}
\caption{The derived abundance patterns for J050622 (upper panel) and
  J052043, obtained with highest S/N spectrum (lower panel).}
\label{fig:abundances1}
\end{center}
\end{figure}

\begin{figure}
\begin{center}
\resizebox{\hsize}{!}{ \includegraphics{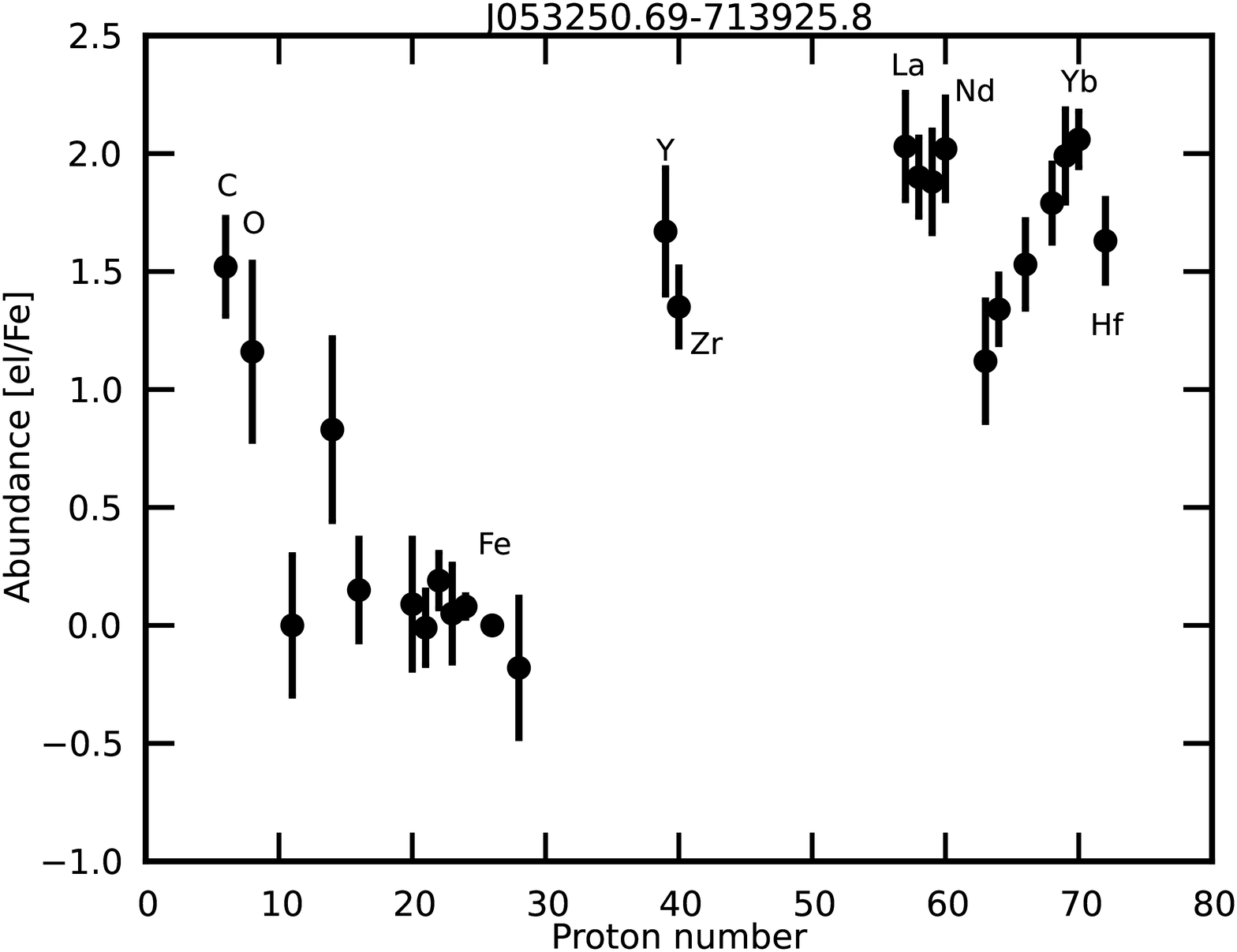}}
\resizebox{\hsize}{!}{ \includegraphics{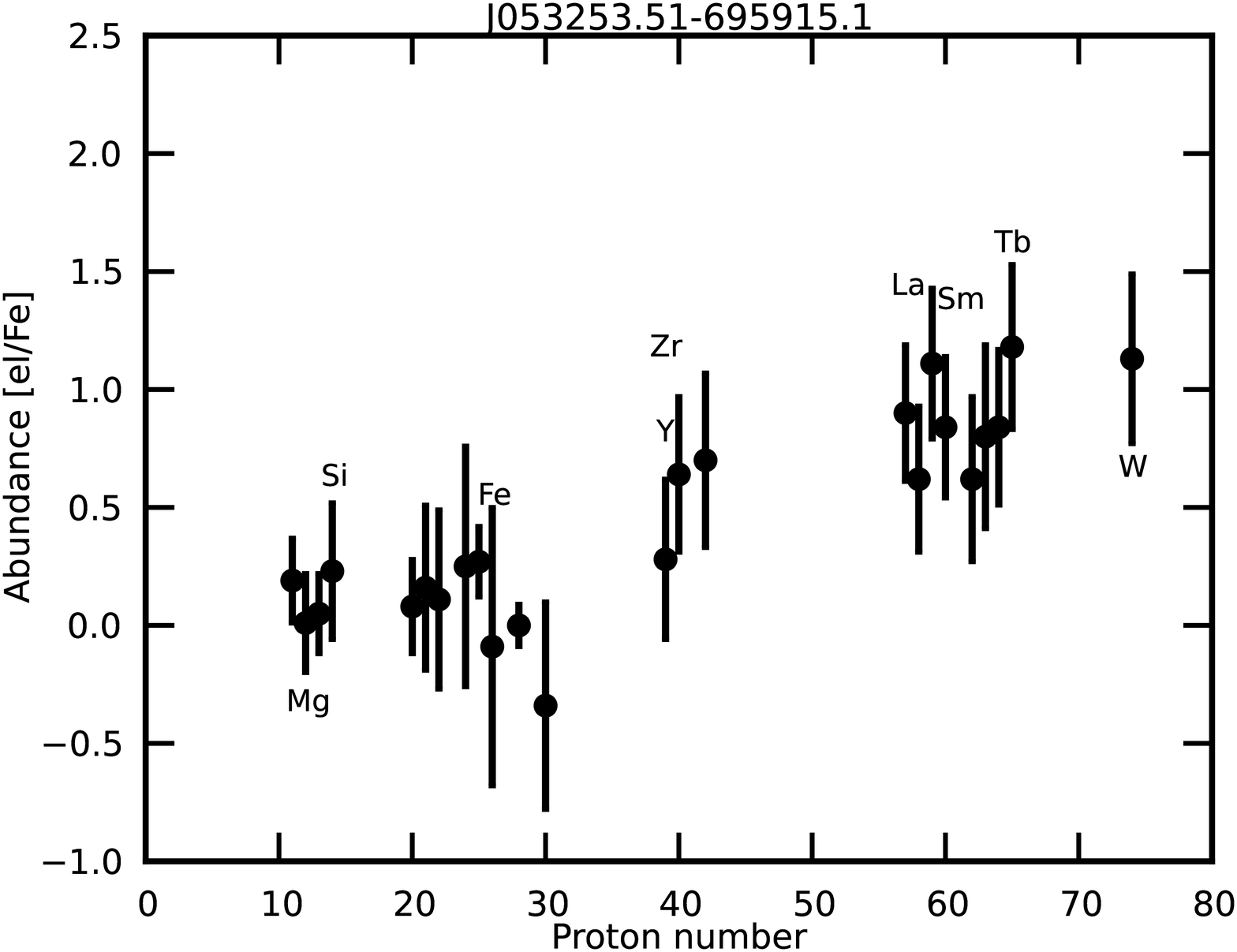}}
\caption{The derived abundance patterns for J053250 (upper panel) and
  J053253, obtained with highest S/N spectrum (lower panel).}
\label{fig:abundances2}
\end{center}
\end{figure}

If less than 5 lines were available, a line-to-line scatter of 0.2~dex
was applied. All results are presented graphically in
Figs.~\ref{fig:abundances1} and \ref{fig:abundances2} . We now highlight the most important
abundance characteristics.

\begin{table}
\vspace{-0.4cm}
\caption{Abundance results for J050632.10-714229.8. \vspace{0.05cm}}             
\label{table:abres1}      
\centering                          
\begin{tabular}{l|rrrrrr|r}        
\hline\hline                 
        & \multicolumn{6}{c|}{J050632} & Sun \\
        & \multicolumn{3}{l}{$T_{\mathrm{eff}} = $ 6750~K} & \multicolumn{3}{r|}{$\xi_t = 3.0$~km/s} &  \\
        & \multicolumn{3}{l}{$\log g = 0.5$} & \multicolumn{3}{r|}{[$\mathrm{Fe}$/$\mathrm{H}$] = -1.22} &  \\
\hline
Ion     & N & $\overline{W_{\lambda}}$\tablefootmark{a} & $\log \epsilon$ & $\sigma_{\mathrm{ltl}}$ & [el/$\mathrm{Fe}$] & $\sigma_{\mathrm{tot}}$ & $\log \epsilon$ \\
 & & (m\AA{}) &  &  &  &  &  \\
\hline
\ion{C}{I} & 21 &            47 &  8.37 &  0.08 &   1.16 &  0.11 &  8.43 \\
\ion{O}{I} &  3 &  \textit{ss}  &  8.18 &       &   0.91 &  0.24 &  8.69 \\
\hline
\ion{Mg}{I} &  1 &           84 &  6.38 &       &  -0.03 &  0.32 &  7.60 \\
\ion{Si}{I} &  1 &            5 &  6.88 &       &   0.59 &  0.32 &  7.51 \\
\ion{S}{I} &  1   & \textit{ss} &  5.93 &       &   0.03 &  0.27 &  7.12 \\
\ion{Ca}{I} & 11 &           23 &  4.98 &  0.10 &  -0.14 &  0.32 &  6.34 \\
\ion{Sc}{II} &  8 &          58 &  1.72 &  0.09 &  -0.20 &  0.1 &  3.15 \\
\ion{Ti}{II} & 18 &          74 &  3.82 &  0.11 &   0.09 &  0.08 &  4.95 \\
\ion{V}{II} &  1 &           74 &  2.48 &       &  -0.23 &  0.22 &  3.93 \\
\ion{Cr}{I} &  1 &           89 &  3.97 &       &  -0.45 &  0.40 &  5.64 \\
\ion{Cr}{II} & 17 &          40 &  4.46 &  0.11 &   0.05 &  0.04 &  5.64 \\
\ion{Fe}{I} & 63 &           36 &  6.09 &  0.18 &  -0.18 &  0.29 &  7.50 \\
\ion{Fe}{II} & 30 &          44 &  6.28 &  0.15 &   0.00 &  0.04 &  7.50 \\
\ion{Ni}{I} &  3 &            8 &  4.62 &  0.03 &  -0.38 &  0.30 &  6.22 \\
\ion{Zn}{I} &  1 &           12 &  3.26 &       &  -0.08 &  0.35 &  4.56 \\
\hline
\ion{Y}{II} &  8 &           88 &  2.53 &  0.05 &   1.54 &  0.11 &  2.21 \\
\ion{Zr}{II} & 12 &          41 &  2.66 &  0.10 &   1.30 &  0.09 &  2.58 \\
\hline
\ion{Ba}{II} &  1 & \textit{ss} &  2.24 &       &   1.28 &  0.34 &  2.18 \\
\ion{La}{II} & 17 &          32 &  1.36 &  0.08 &   1.48 &  0.23 &  1.10 \\
\ion{Ce}{II} & 28 &          48 &  1.68 &  0.10 &   1.33 &  0.21 &  1.58 \\
\ion{Pr}{II} &  8 &          15 &  0.91 &  0.09 &   1.42 &  0.30 &  0.72 \\
\ion{Nd}{II} & 20 &          36 &  1.38 &  0.08 &   1.18 &  0.30 &  1.42 \\
\ion{Sm}{II} &  1 & \textit{ss} &  0.05 &       &   0.29 &  0.33 &  0.96 \\
\ion{Eu}{II} &  2 &          30 & -0.19 &  0.17 &   0.51 &  0.27 &  0.52 \\
\hline
\ion{Gd}{II} & 19 &          22 &  0.81 &  0.11 &   0.96 &  0.16 &  1.07 \\
\ion{Dy}{II} &  5 &          23 &  0.73 &  0.07 &   0.88 &  0.21 &  1.10 \\
\ion{Yb}{II}\tablefootmark{b} &  2 &  5 &  1.31 &  0.12 &   1.70 &  0.15 &  0.84 \\
\ion{Ta}{II}\tablefootmark{b} &  1 & 9 &  0.71 &        &   2.05 &  0.23 & -0.12 \\
\hline                                   
\end{tabular}
\tablefoot{For each element, we list the number of lines involved in the abundance determination, the average equivalent width of these lines, the deduced abundance, the standard deviation on this abundance, and the [el/$\mathrm{Fe}$] value with its corresponding error (see text for details). The solar abundances in the last column are from \cite{asplund09}. If no solar photospheric abundance value was available, as was the case for $\mathrm{Ta}$, $\mathrm{Re}$, and $\mathrm{Hg}$, we took the meteoric equivalent. \\
\tablefoottext{a}{The indication \textit{ss} denotes that the abundance was deduced with the use of spectrum synthesis (see Sect.~\ref{sssec:specsynth}). \\
\tablefoottext{b}{These abundances should be treated with caution as they are based on small features when compared to the noise level, and are upper limits only.}
}}
\end{table}

\begin{table*}
\caption{Same as Table~\ref{table:abres1}, but for the spectra of J052043.86-692341.0 taken in November and December 2008.\vspace{0.05cm}}             
\label{table:abres2}      
\centering                          
\begin{tabular}{l|rrrrrr|r|rrrrrr}        
\hline\hline                 
        & \multicolumn{6}{c|}{J052043\_a} & Sun         & \multicolumn{6}{c}{J052043\_b} \\
        & \multicolumn{3}{l}{$T_{\mathrm{eff}} = $ 5750~K} & \multicolumn{3}{r|}{$\xi_t = 3.0$~km/s} &          & \multicolumn{3}{l}{$T_{\mathrm{eff}} = $ 5750~K} & \multicolumn{3}{r}{$\xi_t = 3.0$~km/s} \\
        & \multicolumn{3}{l}{$\log g = 0.5$} & \multicolumn{3}{r|}{[$\mathrm{Fe}$/$\mathrm{H}$] = -1.24} & 
        & \multicolumn{3}{l}{$\log g = 0.0$} & \multicolumn{3}{r}{[$\mathrm{Fe}$/$\mathrm{H}$] = -1.15} \\
\hline
Ion     & N & $\overline{W_{\lambda}}$\tablefootmark{a} & $\log \epsilon$ & $\sigma_{\mathrm{ltl}}$ & [el/$\mathrm{Fe}$] & $\sigma_{\mathrm{tot}}$ & $\log \epsilon$ & N & $\overline{W_{\lambda}}$\tablefootmark{a} & $\log \epsilon$ & $\sigma_{\mathrm{ltl}}$ & [el/$\mathrm{Fe}$] & $\sigma_{\mathrm{tot}}$ \\
 & & (m\AA{}) &  &  &  &  & &  & (m\AA{}) &  &  &  &   \\
\hline
\ion{C}{I} & 13 &          71  &  8.72 &  0.07 &   1.53 &  0.20 &  8.43  & 11 &          49 &  8.59 &  0.14 &   1.30 &  0.18 \\
\ion{O}{I} &  1 & \textit{ss}  &  8.30 &       &   0.85 &  0.35 &  8.69  &  1 &          15 &  8.40 &       &   0.86 &  0.35 \\
\hline
\ion{Na}{I} &  1 & \textit{ss} &  5.18 &       &   0.18 &  0.35 &  6.24 &    &             &       &       &        &       \\
\ion{Mg}{I} &  1 &          26 &  6.29 &       &  -0.07 &  0.35 &  7.60  &  1 &          36 &  6.54 &       &   0.09 &  0.32 \\
\ion{Si}{I} &  8 &          18 &  6.82 &  0.17 &   0.55 &  0.29 &  7.51  &  4 &          25 &  6.75 &  0.16 &   0.38 &  0.32 \\
\ion{Si}{II} &  2 &         55 &  6.89 &  0.05 &   0.61 &  0.31 &  7.51  &  2 &          16 &  6.86 &  0.09 &   0.50 &  0.31 \\
\ion{S}{I} &  2 &           26 &  6.32 &  0.11 &   0.43 &  0.20 &  7.12  &  2 & \textit{ss} &  6.31 &  0.06 &   0.33 &  0.18 \\
\ion{Ca}{I} & 12 &          55 &  5.20 &  0.10 &   0.10 &  0.32 &  6.34 & 12 &          67 &  5.44 &  0.13 &   0.25 &  0.18 \\
\ion{Ca}{II} &    &            &       &       &        &       & 6.34  &  1 &          10 &  5.46 &       &   0.27 &  0.28 \\
\ion{Sc}{II} &  3 &         80 &  1.96 &  0.13 &   0.04 &  0.16 &  3.15  &  4 &          79 &  1.94 &  0.05 &  -0.06 &  0.16 \\
\ion{Ti}{I}  &    &            &       &       &        &       &  4.95 &  1 &           4 &  3.61 &       &  -0.19 &  0.39 \\
\ion{Ti}{II} &    &            &       &       &        &       &  4.95 &  6 &          94 &  3.90 &  0.12 &   0.09 &  0.10 \\
\ion{Cr}{I} &    &             &       &       &        &       &  5.64 &  2 &          26 &  4.27 &  0.01 &  -0.22 &  0.38 \\
\ion{Cr}{II} & 11 &         56 &  4.54 &  0.09 &   0.13 &  0.05 &  5.64  & 12 &          62 &  4.57 &  0.11 &   0.08 &  0.05 \\
\ion{Fe}{I} & 63 &          59 &  6.16 &  0.19 &  -0.11 &  0.34 &  7.50  & 73 &          60 &  6.25 &  0.16 &  -0.10 &  0.29 \\
\ion{Fe}{II} & 18 &         66 &  6.26 &  0.11 &   0.00 &  0.04 &  7.50  & 23 &          68 &  6.35 &  0.17 &   0.00 &  0.05 \\
\ion{Ni}{I} &  9 &          22 &  4.86 &  0.13 &  -0.12 &  0.33 &  6.22  & 13 &          26 &  4.96 &  0.12 &  -0.11 &  0.27 \\
\ion{Zn}{I} &  1 &          37 &  3.08 &       &  -0.24 &  0.40 &  4.56 \\
\hline
\ion{Y}{II} &  2 & \textit{ss}  &  2.63 &      &   1.86 &  0.21 &  2.21  &  2 &         129 &  2.78 &  0.15 &   1.71 &  0.21 \\
\ion{Zr}{II} &  5 &          46 &  2.81 &  0.16 &   1.47 &  0.12 &  2.58  &  4 &          70 &  2.88 &  0.15 &   1.44 &  0.16 \\
\hline
\ion{La}{II} & 13 &          94 &  1.97 &  0.09 &   2.10 &  0.20 &1.10  &  7 &          89 &  1.80&  0.03 &   1.85 &  0.21 \\
\ion{Ce}{II} &  5 &          66 &  2.03 &  0.13 &   1.68 &  0.16 &  1.58  &  5 &          63 &  2.11 &  0.06 &   1.68 &  0.16 \\
\ion{Pr}{II} &  7 &          79 &  1.62 &  0.12 &   2.14 &  0.21 &  0.72  & 11 &          59 &  1.39 &  0.14 &   1.82 &  0.23 \\
\ion{Nd}{II} & 13 &          85 &  2.14 &  0.10 &   1.96 &  0.21 &  1.42  & 20 &          95 &  2.19 &  0.12 &   1.92 &  0.24 \\
\ion{Sm}{II} &  3 &         127 &  1.38 &  0.10 &   1.66 &  0.24 &  0.96  &  2 &         137 &  1.29 &  0.11 &   1.47 &  0.25 \\
\ion{Eu}{II} &  2 &          74 &  0.52 &  0.09 &   1.24 &  0.21 &  0.52  &  1 &          52 &  0.36 &       &   0.99 &  0.26 \\
\hline
\ion{Gd}{II} &  4 &          17 &  1.04 &  0.13 &   1.21 &  0.17 &1.07  & 9  &          24 &  1.12 &  0.09 &   1.20 &  0.15 \\
\ion{Tb}{II} &  1 &          14 &  0.36 &       &   1.30 &  0.29  &0.30  &  1 &          25 &  0.53 &    &   1.38 &  0.29 \\
\ion{Dy}{II} &  4 &           8 &  1.25 &  0.12 &   1.39 &  0.21 &  1.10  &  1 &          18 &  1.22 &       &   1.27 &  0.30 \\
\ion{Er}{II} &    &             &       &       &        &       &0.92  &    &          13 &  1.27 &  0.08 &   1.50 &  0.16 \\
\ion{Yb}{II} &  3 &          12 &  1.95 &  0.14 &   2.35 &  0.12 &  0.84  &  5 &          15 &  1.80 &  0.13 &   2.11 &  0.07 \\
\ion{Hf}{II} &  1 &          13 &  1.94 &       &   2.33 &  0.21 &0.85  &  1 &          12 &  1.33 &      &   1.63 &  0.24 \\
\hline                                   
\end{tabular}
\end{table*}

\begin{table}
\caption{Same as Table~\ref{table:abres1}, but for J053250.69-713925.8.\vspace{0.05cm}}             
\label{table:abres4}      
\centering                          
\begin{tabular}{l|rrrrrr|r}        
\hline\hline                 
        & \multicolumn{6}{c|}{J053250} & Sun \\
        & \multicolumn{3}{l}{$T_{\mathrm{eff}} = $ 5500~K} & \multicolumn{3}{r|}{$\xi_t = 3.0$~km/s} &  \\
        & \multicolumn{3}{l}{$\log g = 0.0$} & \multicolumn{3}{r|}{[$\mathrm{Fe}$/$\mathrm{H}$] = -1.22} & \\
\hline
Ion     & N & $\overline{W_{\lambda}}$\tablefootmark{a} & $\log \epsilon$ & $\sigma_{\mathrm{ltl}}$ & [el/$\mathrm{Fe}$] & $\sigma_{\mathrm{tot}}$ & $\log \epsilon$ \\
 & & (m\AA{}) &  &  &  &  &  \\
\hline
\ion{C}{I} &  9 &          47 &  8.74 &  0.11 &   1.53 &  0.21 &  8.43 \\
\ion{O}{I} &  1 & \textit{ss}  &  8.35 &       &   0.88 &  0.38 &  8.69 \\
\hline
\ion{Na}{I} &  1 &          23 &  5.01 &       &  -0.01 &  0.30 &  6.24 \\
\ion{Si}{I} &  4 &          35 &  6.84 &  0.12 &   0.55 &  0.25 &  7.51 \\
\ion{Si}{II} &  1 &          29 &  7.15 &       &   0.86 &  0.39 &  7.51 \\
\ion{S}{I} &  1 & \textit{ss} &  6.05 &       &   0.15 &  0.22 &  7.12 \\
\ion{Ca}{I} & 11 &          74 &  5.21 &  0.13 &   0.09 &  0.28 &  6.34 \\
\ion{Sc}{II} &  4 &          88 &  1.92 &  0.05 &  -0.01 &  0.16 &  3.15 \\
\ion{Ti}{II} &  7 &         123 &  3.93 &  0.20 &   0.20 &  0.12 &  4.95 \\
\ion{V}{II} &  1 &         134 &  2.77 &       &   0.06 &  0.21 &  3.93 \\
\ion{Cr}{I} &  2 &          26 &  3.96 &  0.16 &  -0.46 &  0.40 &  5.64 \\
\ion{Cr}{II} & 12 &          52 &  4.50 &  0.12 &   0.08 &  0.05 &  5.64 \\
\ion{Fe}{I} & 58 &          58 &  6.12 &  0.21 &  -0.16 &  0.31 &  7.50 \\
\ion{Fe}{II} & 17 &          70 &  6.28 &  0.10 &   0.00 &  0.03 &  7.50 \\
\ion{Ni}{I} &  9 &          28 &  4.81 &  0.13 &  -0.19 &  0.30 &  6.22 \\
\hline
\ion{Y}{II} &  1 &         118 &  2.66 &       &   1.67 &  0.27 &  2.21 \\
\ion{Zr}{II} &  2 &          67 &  2.74 &  0.03 &   1.35 &  0.17 &  2.58 \\
\hline
\ion{La}{II} &  7 &         106 &  1.91 &  0.13 &   2.03 &  0.23 &  1.10 \\
\ion{Ce}{II} &  8 &          88 &  2.26 &  0.15 &   1.91 &  0.17 &  1.58 \\
\ion{Pr}{II} & 13 &          63 &  1.37 &  0.14 &   1.87 &  0.22 &  0.72 \\
\ion{Nd}{II} &  9 &          91 &  2.22 &  0.13 &   2.03 &  0.22 &  1.42 \\
\ion{Eu}{II} &  1 &          68 &  0.41 &       &   1.11 &  0.26 &  0.52 \\
\hline
\ion{Gd}{II} &  7 &          11 &  1.18 &  0.13 &   1.33 &  0.15 &  1.07 \\
\ion{Dy}{II} &  4 &          26 &  1.39 &  0.10 &   1.52 &  0.19 &  1.10 \\
\ion{Er}{II} &  2 &          12 &  1.49 &  0.09 &   1.79 &  0.17 &  0.92 \\
\ion{Tm}{II}\tablefootmark{b}  &  1 &           4 &  0.87 &       &   1.99 &  0.20 &  0.10 \\
\ion{Yb}{II} &  3 &          10 &  1.68 &  0.05 &   2.06 &  0.12 &  0.84 \\
\ion{Hf}{II}\tablefootmark{b} &  2 &           7 &  1.25 &  0.08 &   1.62 &  0.18 &  0.85 \\
\hline                                   
\end{tabular}
\end{table}

\begin{table}
\caption{Same as Table~\ref{table:abres1}, but for
  J053253.51-695915.1. For this cooler star, \ion{Fe}{I} is the
  dominant ionisation stage.\vspace{0.05cm}}             
\label{table:abres5}      
\centering                          
\begin{tabular}{l|rrrrrr|r}        
\hline\hline                 
        & \multicolumn{6}{c|}{J053253} & Sun \\
        & \multicolumn{3}{l}{$T_{\mathrm{eff}} = $ 4750~K} & \multicolumn{3}{r|}{$\xi_t = 2.5$~km/s} &  \\
        & \multicolumn{3}{l}{$\log g = 2.5$} & \multicolumn{3}{r|}{[$\mathrm{Fe}$/$\mathrm{H}$] = -0.54} & \\ 
\hline
Ion     & N & $\overline{W_{\lambda}}$\tablefootmark{a} & $\log \epsilon$ & $\sigma_{\mathrm{ltl}}$ & [el/$\mathrm{Fe}$] & $\sigma_{\mathrm{tot}}$ & $\log \epsilon$ \\
 & & (m\AA{}) &  &  &  &  &  \\
\hline
\ion{Na}{I} &  2 &          74 &  5.92 &  0.02 &   0.22 &  0.18 &  6.24 \\
\ion{Mg}{I} &  1 &         139 &  7.13 &       &   0.07 &  0.21 &  7.60 \\
\ion{Al}{I} &  2 &          56 &  6.03 &  0.08 &   0.13 &  0.17 &  6.45 \\
\ion{Si}{I} & 15 &          53 &  7.24 &  0.10 &   0.31 &  0.29 &  7.51 \\
\ion{Ca}{I} &  6 &         134 &  5.94 &  0.10 &   0.15 &  0.20 &  6.34 \\
\ion{Sc}{II} &  9 &          82 &  2.85 &  0.09 &   0.24 &  0.35 &  3.15 \\
\ion{Ti}{II} &  8 &         113 &  4.60 &  0.09 &   0.19 &  0.38 &  4.95 \\
\ion{Cr}{I} &  1 &          70 &  5.58 &       &   0.48  &  0.24 &  5.64 \\
\ion{Cr}{II} &  8 &          57 &  5.43 &  0.13 &   0.33 &  0.51 &  5.64 \\
\ion{Mn}{I} &  4 &          87 &  5.23 &  0.18 &   0.34 &  0.15 &  5.43 \\
\ion{Fe}{I} & 60 &          98 &  7.03 &  0.13 &   0.08 &  0.02 &  7.50 \\
\ion{Fe}{II} & 18 &          70 &  6.96 &  0.08 &   0.00 &  0.00 &  7.50 \\
\ion{Ni}{I} & 25 &          88 &  5.75 &  0.12 &   0.07 &  0.09 &  6.22 \\
\ion{Zn}{I} &  1 & \textit{ss} &  3.76 &       &  -0.26 &  0.44 &  4.56 \\
\hline
\ion{Y}{II} &  8 &          74 &  2.03 &  0.14 &   0.36 &  0.34 &  2.21 \\
\ion{Zr}{II} &  6 &          59 &  2.76 &  0.15 &   0.73 &  0.33 &  2.58 \\
\ion{Mo}{I} &  1 &          39 &  2.10 &       &   0.76 &  0.37 &  1.88 \\
\hline
\ion{La}{II} & 12 &          53 &  1.54 &  0.12 &   0.98 &  0.29 &  1.10 \\
\ion{Ce}{II} & 10 &          71 &  1.74 &  0.13 &   0.70 &  0.31 &  1.58 \\
\ion{Pr}{II} &  7 &          37 &  1.37 &  0.12 &   1.19 &  0.32 &  0.72 \\
\ion{Nd}{II} & 22 &          58 &  1.80 &  0.10 &   0.92 &  0.30 &  1.42 \\
\ion{Sm}{II} &  1 &          100 &  1.34 &  0.18 &   0.92 &  0.35 &  0.96 \\
\ion{Eu}{II} &  2 &          48 &  0.86 &  0.06 &   0.88 &  0.39 &  0.52 \\
\hline
\ion{Gd}{II} &  4 &          39 &  1.33 &  0.15 &   0.81 &  0.33 &  1.07 \\
\ion{Tb}{II} &  2 &          21 &  0.98 &  0.06 &   1.22 &  0.35 &  0.30 \\
\ion{W}{I} &  1 &            47 &  1.48 &  0.20 &   1.17 &  0.36 &  0.85 \\
\hline                                   
\end{tabular}
\end{table}

\subsection{Metallicity}

J050632, J052043, and J053250 are strongly metal deficient with a
[$\mathrm{Fe}$/$\mathrm{H}$] of $-1$.2 (see
Table~\ref{table:abratios}).  J053253 is only mildly metal deficient
with $[\mathrm{Fe}/\mathrm{H}] = -0.5$. Overall, the metallicities of the stars in
the LMC range from [$\mathrm{Fe}$/$\mathrm{H}$] $= -2.0$ up to
[$\mathrm{Fe}$/$\mathrm{H}$] $= -0.3$, with most objects having
[$\mathrm{Fe}$/$\mathrm{H}$] $\sim -0.5$ \citep{geisler09}.
The fact that three of the four programme stars display metallicities on the
low-end of the detected range, indicates 
that these are old, evolved objects with low initial masses.

For each object, the other iron peak elements follow this
deficiency but the number of detected useable lines is much
lower. The relative abundances of
nickel in J050632 and of zinc in J052043 and J053253 are somewhat
low.

The non-LTE ionisation effect in J050632, J052043, and J053250 can be seen in the
abundances of the other elements of the iron peak if results of more
than one ionisation state are present. Both titanium and chromium show
this bias, with derived abundances on the basis of neutral lines being significantly lower
than the abundances based on ionised lines. 
This non-LTE effect is not seen is silicon and calcium where
abundances based on differen ions are in accordance. 
Only in
the spectrum of J053250 does the silicon abundances deviate from this
trend with \ion{Si}{I} having an abundance of 6.85, while it is 7.12
for \ion{Si}{II}. In the abundance pattern of J053253, the non-LTE
effect is not witnessed, and the obtained abundance of \ion{Cr}{I} is
even higher than that of \ion{Cr}{II}.

\subsection{C/O-ratio}

All objects for which the determination of the carbon abundance was
possible, are
clearly enriched, with values of
$[\mathrm{C}/\mathrm{Fe}]$ ranging from 1.5 in J050632 to 2.5 in
J053250. Surprisingly, large enhancement in oxygen
is also observed in these objects with $[\mathrm{O}/\mathrm{Fe}]$ around
0.8 in J052043, J050632 and J05325. The $\mathrm{C}/\mathrm{O}$ number
ratios remain relatively moderate as a consequence of this large
oxygen enhancement, with values of 2.6 to 1.5 (see Table~\ref{table:abratios}). No conclusions on
both the carbon and oxygen abundance of  J053253 are possible as at the
low effective temperature of this object, no lines of these elements
are visible in the optical wavelength domain.


\subsection{$\alpha$-elements}

The mean of the [el/$\mathrm{Fe}$] ratios for the $\alpha$-elements
$\mathrm{Mg}$, $\mathrm{Si}$, $\mathrm{S}$, $\mathrm{Ca}$, and
$\mathrm{Ti}$, falls between 0.1 and 0.3 for
all objects (see Table~\ref{table:abratios}). This is slightly
deficient when compared to stars of the same metallicities in the
Galaxy, but consistent with the $\alpha$-enhancement expected for LMC
stars at their respective metallicities \citep{pompeia08}. 
Because the $\alpha$-elements are
predicted to be mainly produced in high-mass type~II
supernovae, \cite{pompeia08} interpret the moderate
$\alpha$-enhancement at low metallicites in the LMC as indicative of a stronger influence of type~Ia supernovae.
This also accords with the more extended star formation history in the LMC
than in the Galaxy.

\subsection{s-process elements}

In general, the s-process elements observed in evolved stars are
subdivided into three groups, depending on the number of neutrons in
their nuclei. The light s-process elements have neutron numbers around
the magic number~50 ($\mathrm{Y}$, $\mathrm{Zr}$, \ldots), and the
neutron numbers of the heavy s-process elements are distributed around
the magic number~82 ($\mathrm{Ba}$, $\mathrm{La}$,
$\mathrm{Ce}$, $\mathrm{Pr}$, $\mathrm{Nd}$, $\mathrm{Sm}$,
\ldots). The third and final group corresponds to the double magic
state of neutron number~126 and consists of lead only
($^{208}\mathrm{Pb}$ contains 126~neutrons and 82~protons), as it is
seen as the end product of the s-process nucleosynthesis.

From Table~\ref{table:abratios} and Figs.~\ref{fig:abundances1} and ~\ref{fig:abundances2} it is clear
that all objects are strongly enriched in s-process elements, which
confirms the post third dredge-up status of our stars. The enrichment in s-process
elements is the lowest in J053253, where $[\mathrm{s}/\mathrm{Fe}] =
0.85$. J053250 and J052043 are s-process enhanced at almost the level of
IRAS06530-0213 \citep{reyniers04}, and IRAS
05341+0852 \citep{vanwinckel00}, which are, with respectively
$[\mathrm{s}/\mathrm{Fe}] = 2.1$ and $[\mathrm{s}/\mathrm{Fe}] = 2.2$,
some of the most s-process enriched intrinsic Galactic objects known to date.

\begin{table*}
\caption{The $\mathrm{C}/\mathrm{O}$ number ratio, metallicity, enhancement in $\alpha$-elements and the s-process indexes for the stars in our sample.}             
\label{table:abratios}      
\centering                          
\begin{tabular}{lrrrrr}        
\hline\hline                 
Object     & \multicolumn{1}{c}{$\mathrm{C}/\mathrm{O}$} & \multicolumn{1}{c}{$[\mathrm{Fe}/\mathrm{H}]$} & \multicolumn{1}{c}{$[\alpha/\mathrm{Fe}]$} & \multicolumn{1}{c}{$[\mathrm{s}/\mathrm{Fe}]$} & \multicolumn{1}{c}{$[\mathrm{s}/\mathrm{Fe}]_{\mathrm{La, Nd}}$} \\
\hline
J050632    & 1.5 $\pm$ 0.3 & -1.22 $\pm$ 0.16 & 0.11 $\pm$ 0.09 & 1.17 $\pm$ 0.09 & 1.38 $\pm$ 0.07 \\
J052043\_a & 2.6 $\pm$ 1.3 & -1.24 $\pm$ 0.12 & 0.26 $\pm$ 0.08 & 1.87 $\pm$ 0.06 & 1.84 $\pm$ 0.07 \\
J052043\_b & 1.6 $\pm$ 0.9 & -1.15 $\pm$ 0.17 & 0.23 $\pm$ 0.07 & 1.75 $\pm$ 0.06 & 1.74 $\pm$ 0.07 \\
J053250    & 2.5 $\pm$ 0.7 & -1.22 $\pm$ 0.11 & 0.26 $\pm$ 0.07 & 1.83 $\pm$ 0.07 & 1.78 $\pm$ 0.07 \\
J053253    &               & -0.54 $\pm$ 0.09 & 0.18 $\pm$ 0.07 & 0.85 $\pm$ 0.07 & 0.75 $\pm$ 0.08 \\
\hline
\hline
Object & \multicolumn{1}{c}{$[\mathrm{ls}/\mathrm{Fe}]$} & \multicolumn{1}{c}{$[\mathrm{hs}/\mathrm{Fe}]$} & \multicolumn{1}{c}{$[\mathrm{hs}/\mathrm{Fe}]_{\mathrm{La, Nd}}$} & \multicolumn{1}{c}{$[\mathrm{hs}/\mathrm{ls}]$} & \multicolumn{1}{c}{$[\mathrm{hs}/\mathrm{ls}]_{\mathrm{La, Nd}}$} \\
\hline
J050632    & 1.42 $\pm$ 0.05 & 1.06 $\pm$ 0.14 & 1.33 $\pm$ 0.14 & -0.36 $\pm$ 0.15 & -0.09 $\pm$ 0.15 \\
J052043\_a & 1.65 $\pm$ 0.10 & 1.95 $\pm$ 0.07 & 1.84 $\pm$ 0.07 &  0.30 $\pm$ 0.13 &  0.39 $\pm$ 0.13 \\
J052043\_b & 1.58 $\pm$ 0.10 & 1.79 $\pm$ 0.08 & 1.91 $\pm$ 0.09 &  0.21 $\pm$ 0.13 &  0.34 $\pm$ 0.14 \\
J053250    & 1.53 $\pm$ 0.11 & 1.96 $\pm$ 0.08 & 2.03 $\pm$ 0.09 &  0.43 $\pm$ 0.14 &  0.50 $\pm$ 0.14 \\
J053253    & 0.55 $\pm$ 0.10 & 0.94 $\pm$ 0.10 & 0.95 $\pm$ 0.13 &  0.39 $\pm$ 0.14 &  0.41 $\pm$ 0.16 \\
\hline 
\end{tabular}
\end{table*}


%
%

\section{Luminosities and initial masses}\label{sec:lum_initmass}

With the best atmosphere models as followed from the abundance
analysis, we are now able to make a better estimate of the
luminosities, and consequently initial masses of the stars in our
sample.

\subsection{Luminosity}

\begin{figure}
\resizebox{\hsize}{!}{
\includegraphics{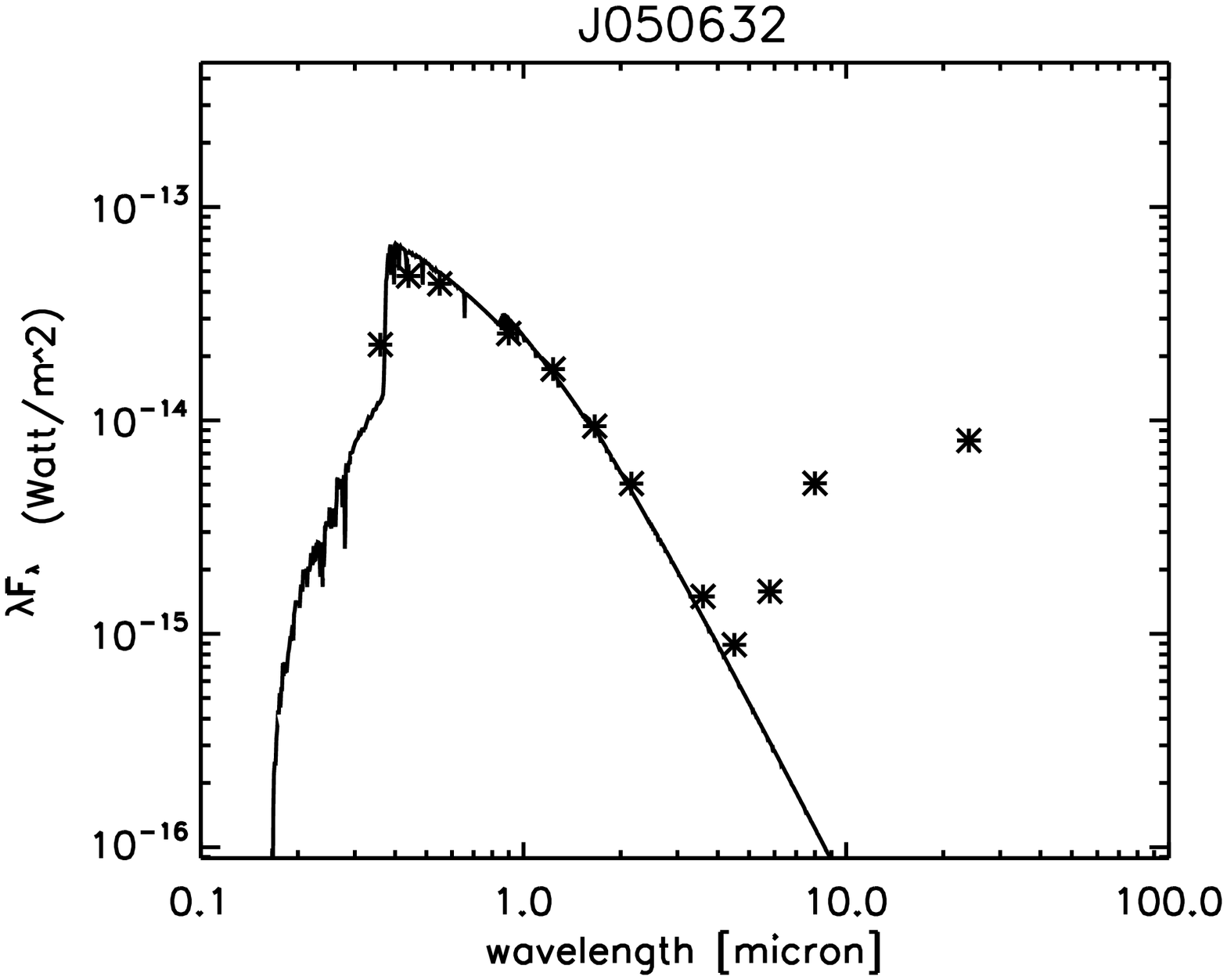}
\includegraphics{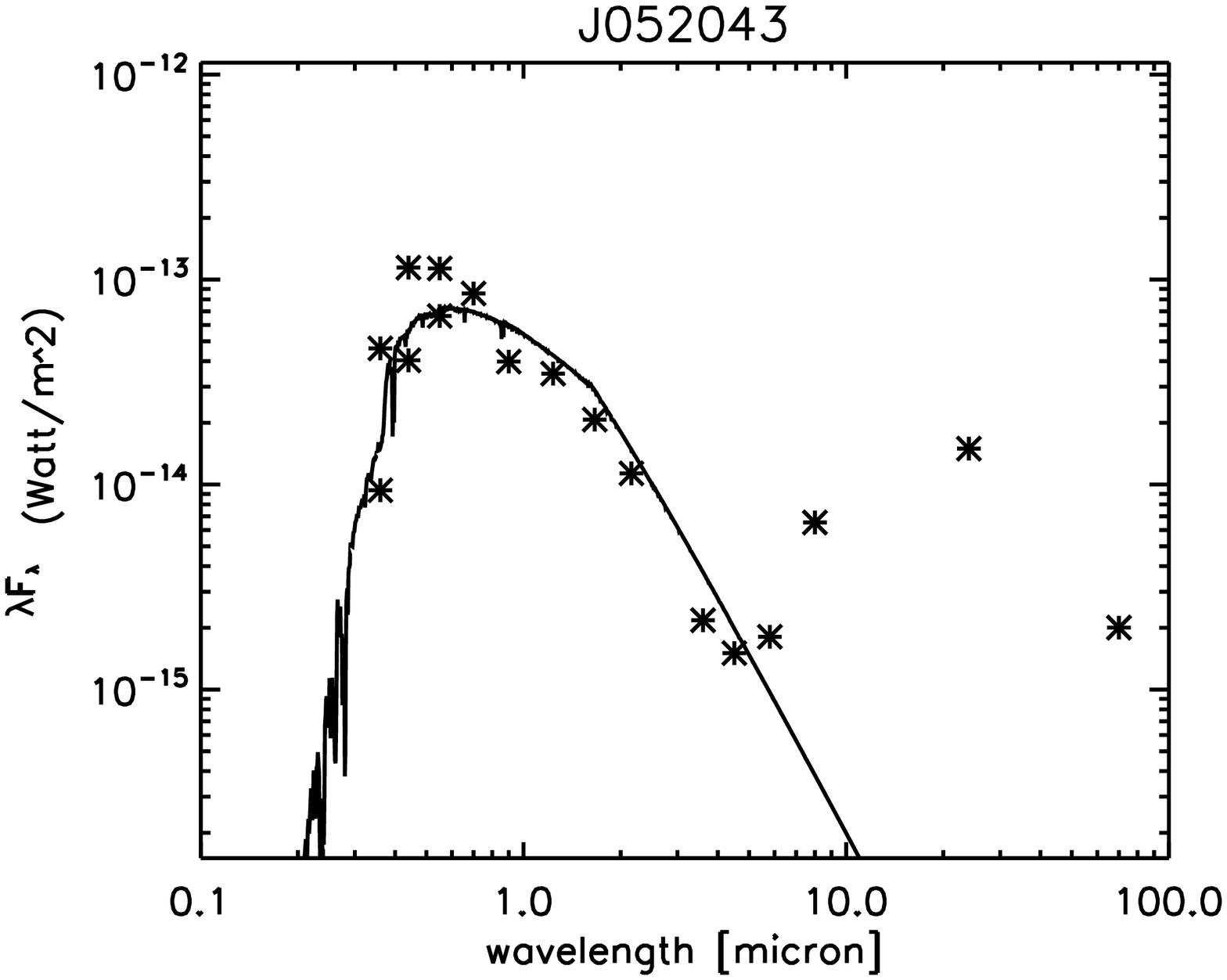}
}
\resizebox{\hsize}{!}{
\includegraphics{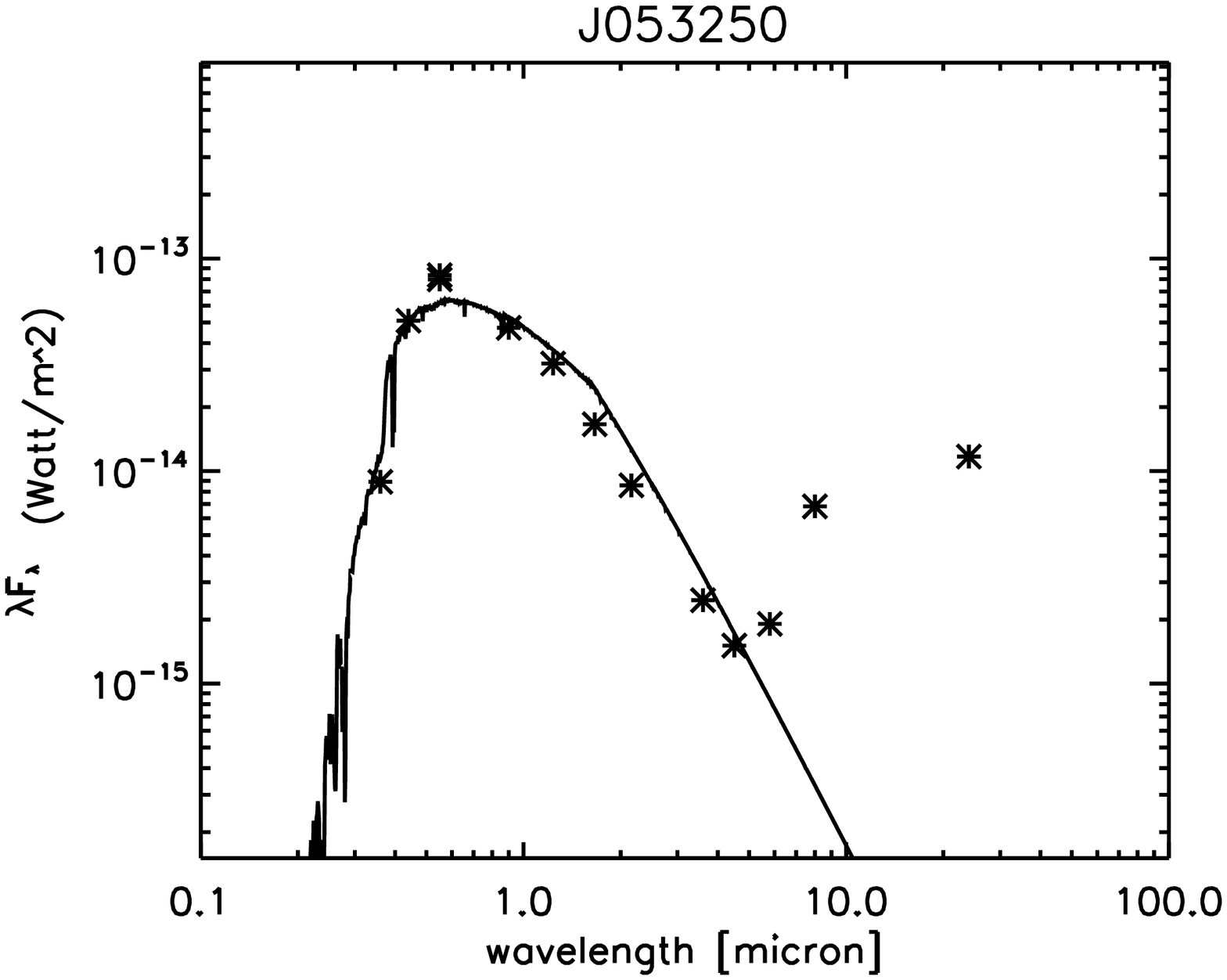}
\includegraphics{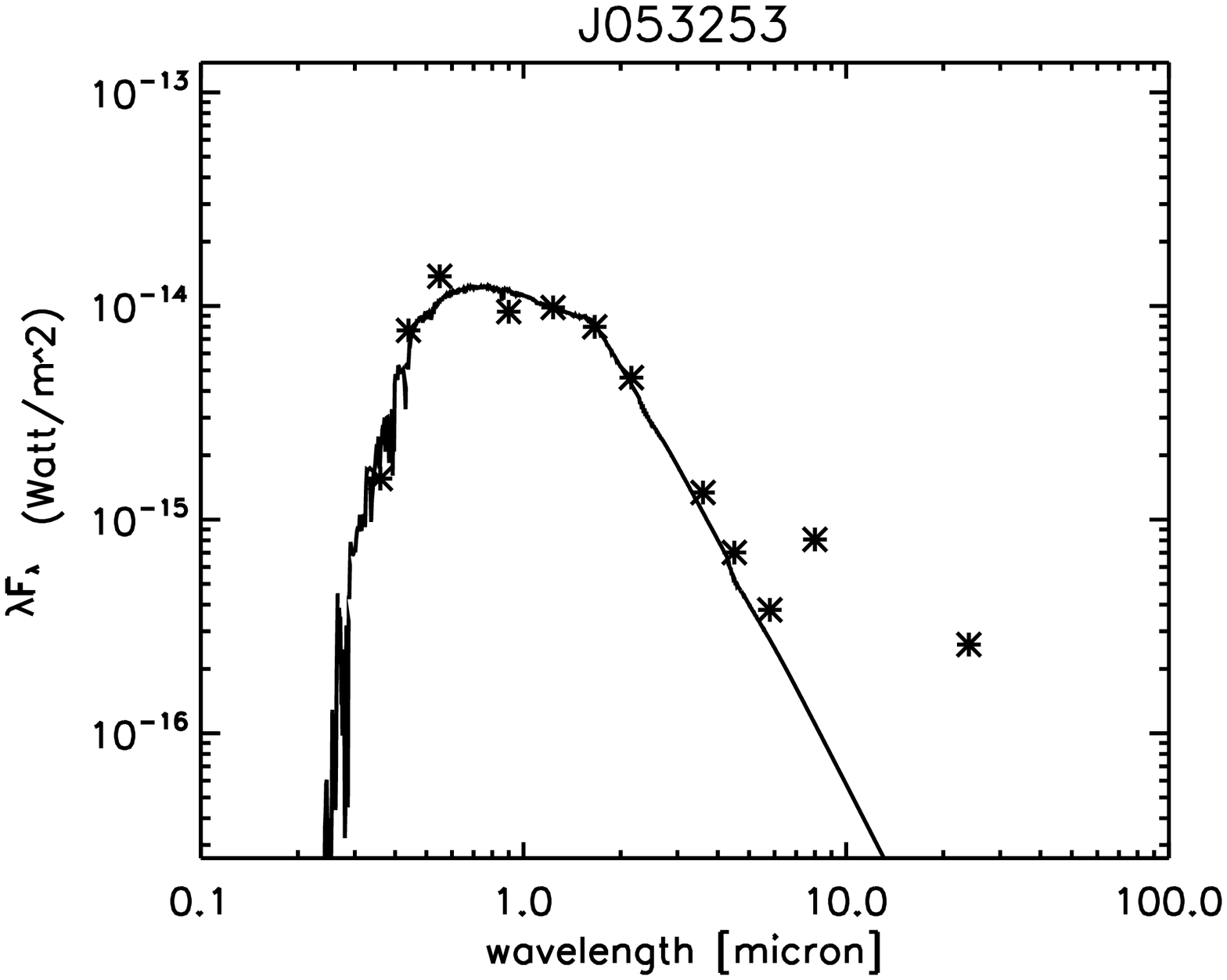}
}
\caption[Final SEDs of the stars in our program.]{Final SEDs of the
  stars in our program. We fitted the dereddened photometry to the
  rescaled atmosphere models, the parameters of which are based on the abundance analysis.}
\label{fig:SEDnew}
\end{figure}

In Fig.~\ref{fig:SEDnew} we show the SEDs of the different stars,
where the appropriate atmosphere model (see Sect.~\ref{ssec:atmod})
was fitted to the dereddened photometry. The E(B-V) was determined by
minimising the difference between the scaled model and the dereddened
photometric data, excluding the infrared where thermal emission of
dust contributes to the total fluxmeasurement. Errors in the luminosity and
the reddening are dominated by the uncertainty in the parameters of
the model atmosphere which are more important than the uncertainties
of the photometric data points themselves. We therefore ran our SED
fitting routine for all input models within the uncertainties of the
spectral analyses.

\begin{table}
\caption[Comparison of luminosity and reddening values obtained with our different computation methods.]{Comparison of luminosity and reddening values obtained with our different computation methods.}             
\label{table:complum}      
\centering                          
\begin{tabular}{cccccc}        
\hline\hline                 
Object  & \multicolumn{3}{c}{Final results} & SED integral. \\
        & $T_{\mathrm{eff}}$ & $L$            & E(B-V)          & $L$  \\
        &                (K) & ($L_{\odot}$)  &                 & ($L_{\odot}$)  \\
\hline
J050632 &     $6750 \pm 250$ & $5400 \pm 700$  &$0.05 \pm 0.06$ & 5570 \\
J052043 &     $5750 \pm 250$ & $8700 \pm 1000$ & $0.43 \pm 0.07$ & 5730 \\ 
J053250 &     $5500 \pm 250$ & $6500 \pm 1000$  & $0.50 \pm 0.08$ & 4700 \\
J053253 &     $4750 \pm 250$ & $1400 \pm 300$  & $0.21 \pm 0.12$ & 1150 \\
\hline
\end{tabular}
\end{table}

In Table~\ref{table:complum} we compare our luminisity estimates of the different
program stars, with those obtained by integrating the raw photometric
data over the full SED (in the table given in the column raw SED integral). The luminosity
based on integral of the raw data holds a good estimate of
the luminosity based on the model atmosphere fit. This indicates that
the circumstellar shell is spherically symmetric,
and that the reddening caused by the interstellar medium is
negligible. With these assumptions, the interstellar reddening can be
neglected, while the extinction caused by the circumstellar shell is
accounted for by the luminosity of the IR excess.

In \cite{vanaarle11},  we used the spectral type to determine
a range of possible effective temperatures and concluded
simultaneously on the best reddening factor and effective temperature
by fitting atmosphere models to the dereddened photometry. We used,
however, only models with metallicity -0.3, which is too high for three of the
four stars we discuss here in detail. These objects are intrinsically bluer
than the used models, which leads to an overestimation of the effective temperature,
and consequently a too high reddening and luminosity. This lead in
\cite{vanaarle11} to an overestimate of the luminosity and the
integral over the raw photometry is a better estimate of the true
luminosity. In Kamath et al. (2013) we report on our low-resolution
spectral fitting routine to obtain more accurate model photospheres on
the basis of low-resolution spectra.


\subsection{Initial mass}\label{ssec:inmass}

\begin{figure}
\begin{center}
\resizebox{9cm}{!}{\includegraphics{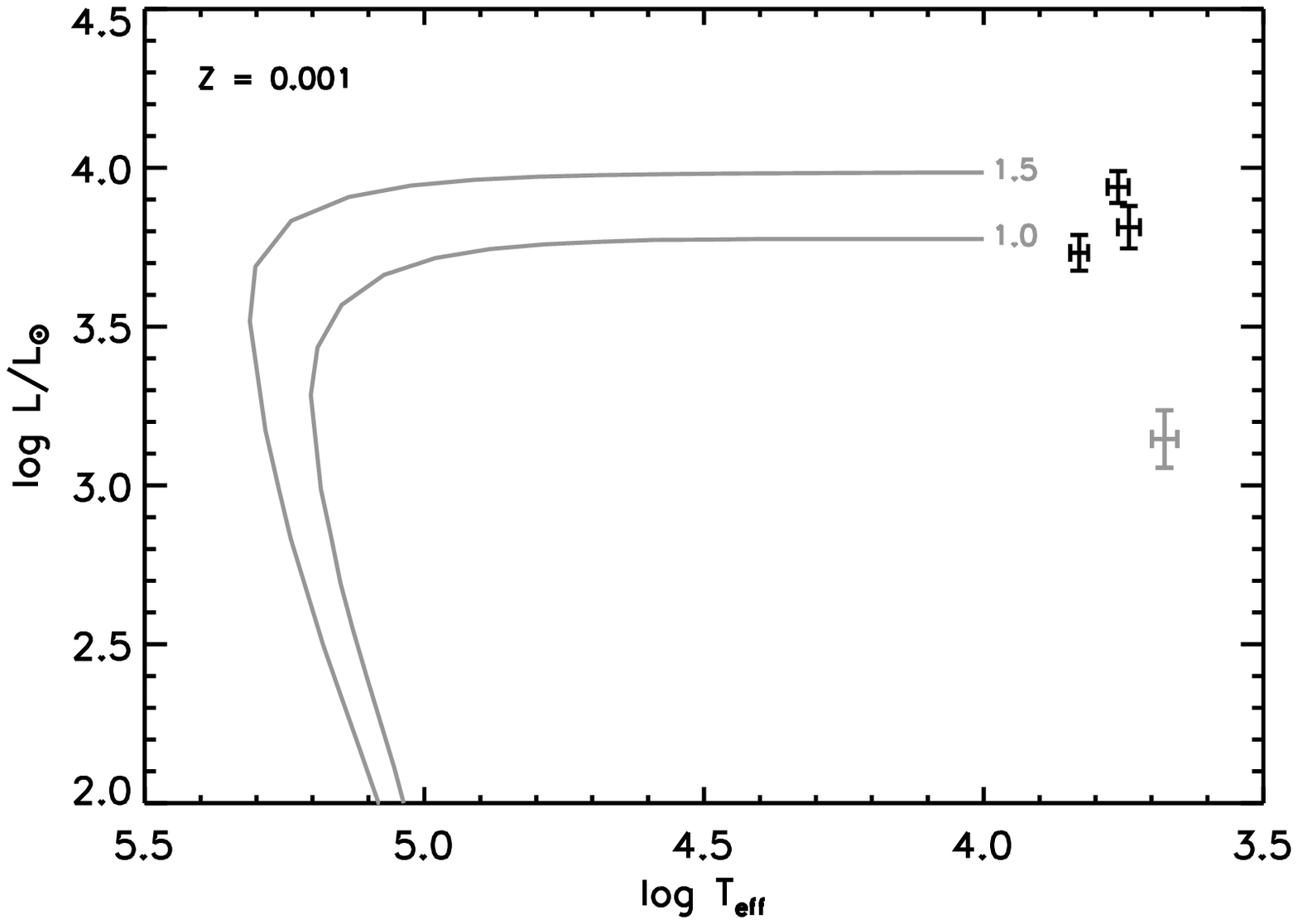}}
\resizebox{9cm}{!}{\includegraphics{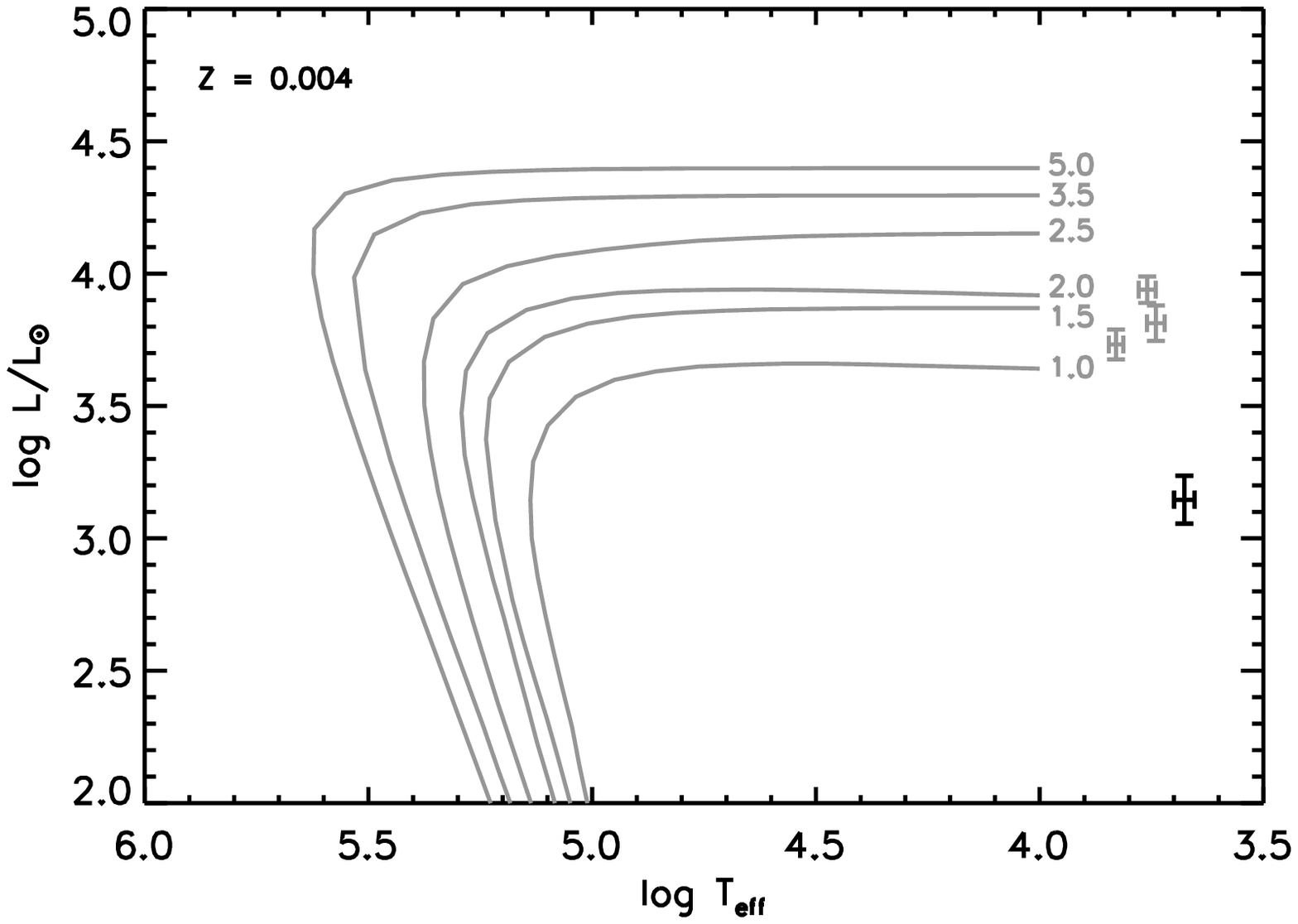}}
\end{center}
\caption[The position of our stars in the HR diagram.]{ The position
  of our stars in the HR diagram. The
  displayed evolutionary tracks are from \cite{vassiliadis94}, with
  metallicities of $Z = 0.001$ (upper panel), and $Z = 0.004$ (lower
  panel), and are labelled with their initial mass in $M_{\odot}$. The
  location of each star is shown in black in the appropriate diagram
  and in grey in the other to compare the position of the different
  data points.}
\label{fig:hrdiagr}
\end{figure}

%
%

The more accurate values of effective temperature and luminosity allow
us to estimate the initial mass of the stars in our sample. In
Fig.~\ref{fig:hrdiagr} the position of the objects in the HR~diagram
is shown, together with the evolutionary tracks from
\cite{vassiliadis94}. The tracks in the upper plot have a metallicity
of $Z = 0.001$ which is the lowest value in the model grid of
\cite{vassiliadis94} but is still somewhat higher than what we find
for J050632 ($Z = 0.0007$), J052043 ($Z = 0.0008$), and J053250 ($Z =
0.0007$). The tracks in the lower plot correspond to $Z = 0.004$,
which is the grid value that approaches the metallicity of J053253 ($Z
= 0.0040$) the best. We converted the metallicities of the stars in
our program to $Z$-values by computing this parameter with its
definition, under the assumptions that the Fe-abundance remained the
same and that all initial abundances, except for those of hydrogen and helium,
scale with the iron abundance.

From Fig.~\ref{fig:hrdiagr} it follows that the initial masses of the
stars in our sample are all smaller than about 1.5~$M_{\odot}$, under
the assumption that these objects follow the single star evolutionary
tracks. J052043 appears to be the most massive, followed by J053250
and J050632, although the difference is small.  J053253 is located
below the evolutionary tracks, which would indicate an significantly
lower initial mass. This is difficult to understand as objects with
masses below ~0.8 M$_{\odot}$ have main sequence life times that are
larger that the Hubble time. We conclude that J053253 is likely not
evolving along the single-star evolutionary tracks depicted here.
The evolutionary status of this object is discussed
further in Section 8.

\section{Neutron exposure}\label{sec:neutrexp}

Four observational indexes are traditionally defined to describe the
s-process overabundances as well as the s-process distribution:
$[\mathrm{s}/\mathrm{Fe}]$, $[\mathrm{ls}/\mathrm{Fe}]$,
$[\mathrm{hs}/\mathrm{Fe}]$, and $[\mathrm{hs}/\mathrm{ls}]$. The
specific elements taken into account when calculating these indexes,
vary from author to author, and mainly depend on which abundances
could be determined and the reliability of these results. To be
consistent with earlier publications \citep[e.g.,][]{reyniers04}, we
follow the suggestion from \cite{busso95} and define the ls-index as
the mean of the relative abundances of $\mathrm{Y}$, and
$\mathrm{Zr}$, and the hs-index as the mean of $\mathrm{Ba}$,
$\mathrm{La}$, $\mathrm{Nd}$, and $\mathrm{Sm}$. Consequently,
$[\mathrm{s}/\mathrm{Fe}]$ is the mean of the  six elements and
$[\mathrm{hs}/\mathrm{ls}] = [\mathrm{hs}/\mathrm{Fe}] -
[\mathrm{ls}/\mathrm{Fe}]$.

The s-process indexes of our sample stars are listed in
Table~\ref{table:abratios}. Missing abundances of specific  s-process
elements need to be determined, however. For this we scaled the
abundance distribution prediction of the most appropriate AGB model (see Section~\ref{sec:agbmod}) to
the nearest element in atomic mass with a measured abundance.
We then assume that the object displays the same abundance ratio for
elements nearest in atomic mass. The $\mathrm{Ba}$ abundance of J052043\_a,
J052043\_b, J053250, and J053253, and the $\mathrm{Sm}$ abundance of
J053250 had to be estimated this way. The four heavy s-process elements all originate from the same iron seeds
and we assume that their relative abundances only depend on their cross-sections for
neutron capture. Leaving the missing abundances out is not a reliable
alternative as due to the odd-even effect, the
overabundances of these four elements are not the same. Alternative
indexes of comparable quality but based on the abundances of La, and Nd only 
\citep{reyniers04} can also be obtained,
and we list these for comparison.

\begin{figure}
\begin{center}
\resizebox{9cm}{!}{\includegraphics{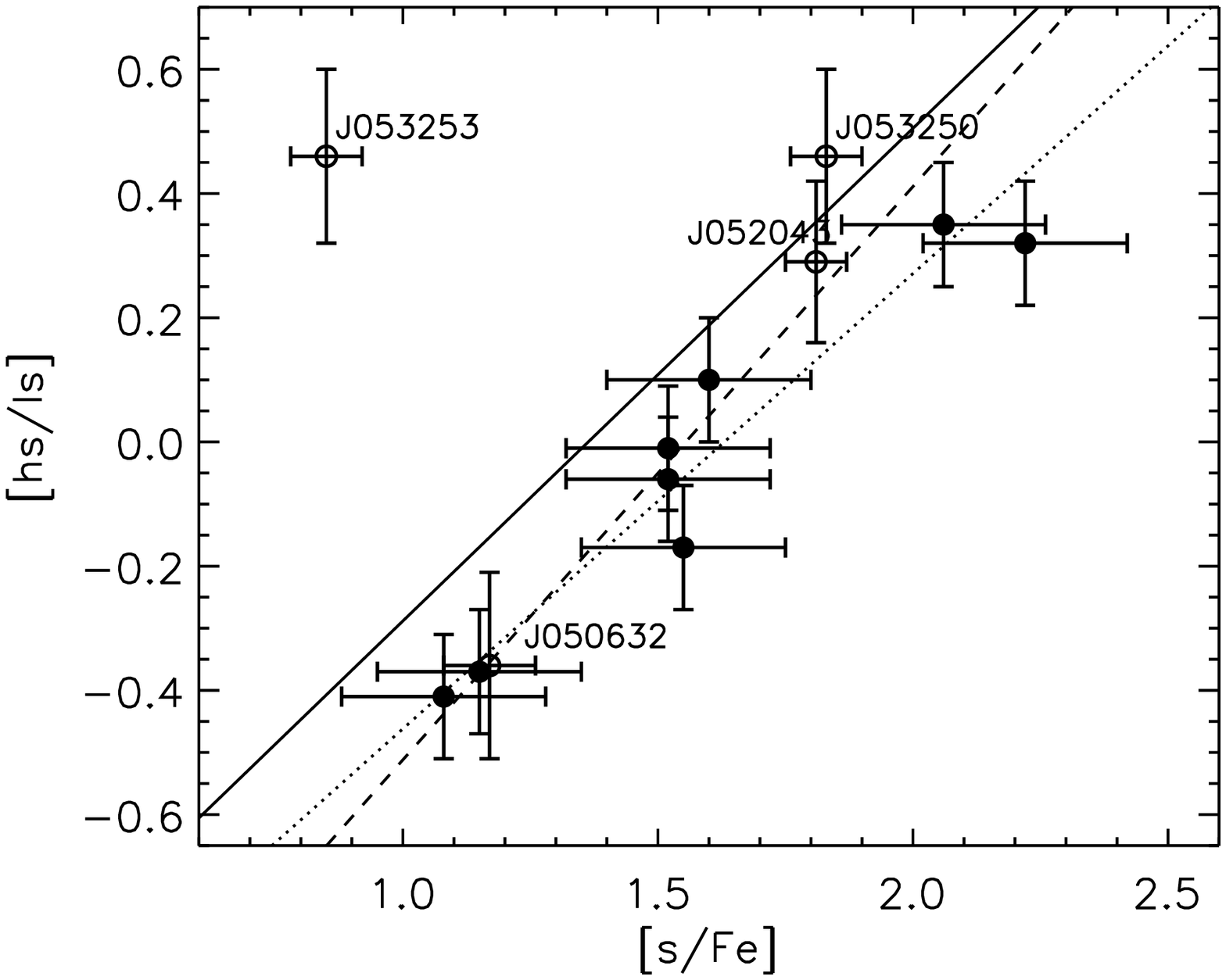}}
\end{center}
\caption[{The correlation between the total enrichment in s-process
  elements and the [hs/ls] index.}]{The correlation between the total
  enrichment in s-process elements and the [hs/ls] index. The objects
  of this paper are named. The others are the 21~$\mu$m sources from \cite{vanwinckel00}, and the two enriched objects from \cite{reyniers04}. The full line gives the least-squares fit to all data points, the dashed line to all data points except for J053253, and the dotted line, which is given for comparison, to all older data points.}
\label{fig:sfevshsls}
\end{figure}

Observationally, it is well known that there is a strong correlation
between the [s/$\mathrm{Fe}$] and [hs/ls] indexes
\citep[e.g.,][]{vanwinckel00, reyniers04}, with objects with a higher
[s/$\mathrm{Fe}$] index showing also a higher [hs/ls] ratio. The
[s/$\mathrm{Fe}$] index forms an indication of the third dredge-up
efficiency, although this parameter is also influenced by the mass
loss history of the star and the envelope mass at dredge-ups. The [hs/ls] index
represents the effectiveness of the neutron irradiation, which is
related to the ratio of the number of neutrons available to the iron
seed nuclei. Considering that the stars in our sample are of similar
metallicity, this parameter would be determined by the abundance
difference between $^{13}\mathrm{C}$, the main neutron seed, and
$^{14}\mathrm{N}$, the main neutron poison in the stellar
interior \citep[e.g.][]{herwig05}. This difference, however, depends on many unconstrained
parameters. The neutron production in AGB models is fuelled by the
injection of protons into the intershell, which creates the
$^{13}\mathrm{C}$ pocket through $^{12}\mathrm{C}(\mathrm{p},
\gamma)^{13}\mathrm{N}(\beta^+)^{13}\mathrm{C}$, and allows the
neutron producing reaction $^{13}\mathrm{C}(\alpha,
\mathrm{n})^{16}\mathrm{O}$ to take place. The correlation therefore
indicates that with increasing dredge-up efficiency, also the
dredge-in of protons will increase as they are the primary cause of a
higher neutron irradiation.

The correlation between the strength of the neutron irradiation, and
the efficiency of the third dredge-up is graphically depicted in
Fig.~\ref{fig:sfevshsls}. Three of the four stars in our sample
further confirm this assertion, but in J053253 the relatively high
neutron irradiation only led to a mild enhancement in s-process
elements. A simple, linear least-squares fit to all data points gives
$[\mathrm{hs}/\mathrm{ls}] = 0.55 \times [\mathrm{s}/\mathrm{Fe}] -
0.78$, and a correlation coefficient of 0.52. If we ignore J053253,
this relation becomes $[\mathrm{hs}/\mathrm{ls}] = 0.92 \times
[\mathrm{s}/\mathrm{Fe}] - 1.43$, with a correlation coefficient of
0.92. This is comparable to the correlation coefficient of 0.96 that
corresponds to the original data \citep{reyniers04}.

 

\begin{figure}
\begin{center}
\resizebox{9cm}{!}{\includegraphics{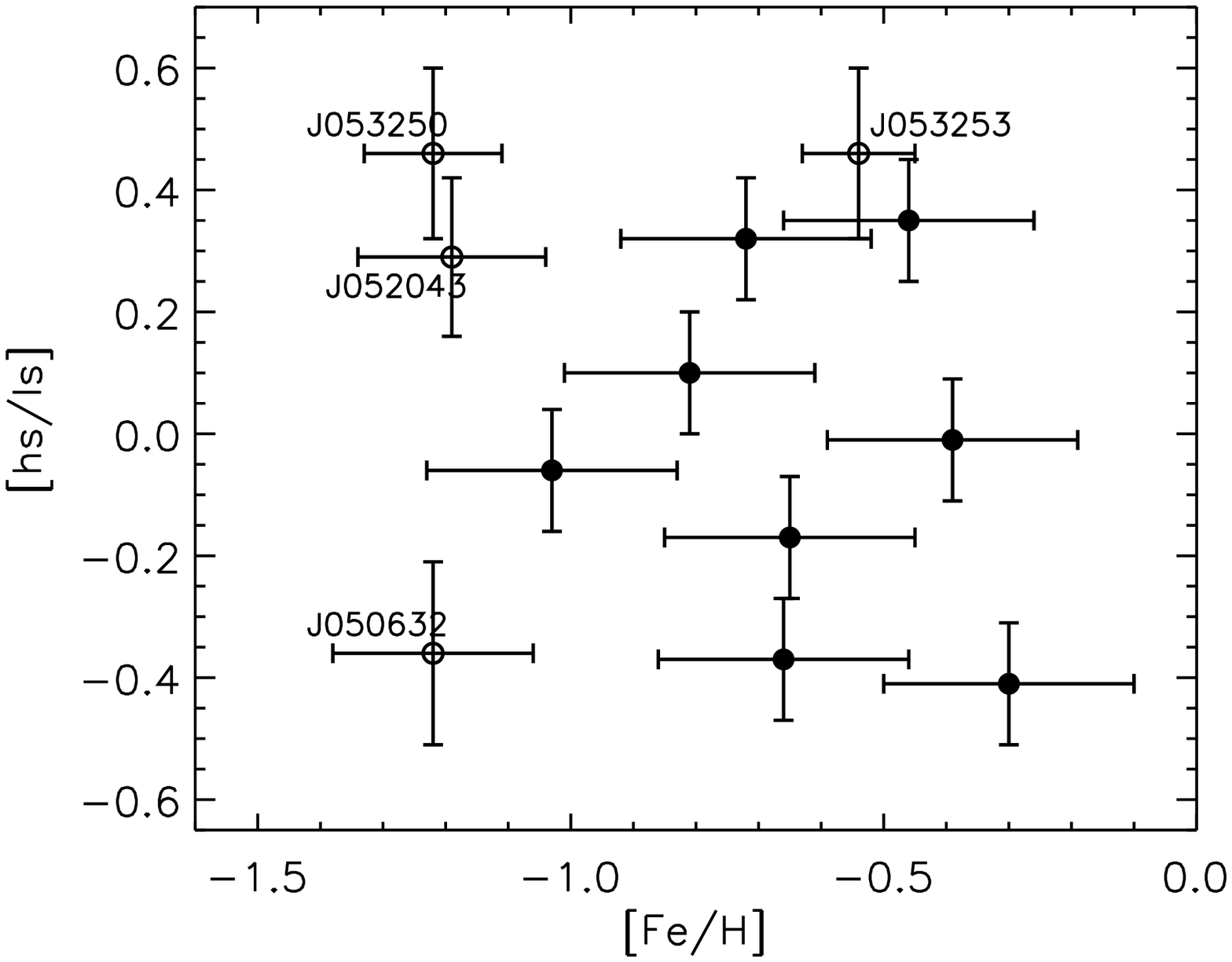}}
\end{center}
\caption[{The [hs/ls] index as a function of the metallicity.}]{The [hs/ls] index as a function of the metallicity. The symbols are the same as in Fig.~\ref{fig:sfevshsls}. No correlation is found between the metallicity and the [hs/ls] index in this range of metallicities.}
\label{fig:fehvshsls}
\end{figure}

The [hs/ls] index is expected to increase with decreasing metallicity
as a consequence of the larger number of neutrons per iron seed
nucleus. This assumes, however, that the diffusion of protons into the
$\mathrm{C}$-rich intershell itself is independent of the metallicity
or parameters linked to this
metallicity. 
Only a weak correlation with a broad intrinsic spread is observed
between the neutron irradiation and the metallicity
\citep[e.g.,][]{vanwinckel00,reyniers04}, which indicates that
other stellar parameters also determine the internal nucleosynthesis during
the AGB evolution, and we are dealing with a large spread in
$^{13}\mathrm{C}$ pocket efficiencies in highly similar objects.

With their range of metallicities from -0.54 to -1.22, our objects
offer the possibility to extend this relation to lower values of
[$\mathrm{Fe}$/$\mathrm{H}$]. In Fig.~\ref{fig:fehvshsls} we added the
current sample of post-AGB stars to the objects from
\cite{vanwinckel00} and \cite{reyniers04} in a plot of [hs/ls] versus
[$\mathrm{Fe}$/$\mathrm{H}$]. 
Our results confirm that there is no correlation between metallicity
and the [hs/ls] index in this range of metallicities.

\section{Is J053253 an extrinsic object?}\label{sec:J053253extr}

Objects that are enhanced in s-process elements, can have reached this
status in two different ways. In the case of intrinsic post-AGB stars,
the surface composition became s-process enriched by dredge-ups of
nucleosynthetic products created in the stellar interior. In extrinsic
objects, the enrichment is an effect of mass transfer of s-process
elements from a further evolved companion to the star at hand. The
former AGB star is now typically a cool white dwarf. Therefore, it is
possible that the enhanced object itself is not yet a post-AGB star.

The strange position of J053253 in Figs.~\ref{fig:hrdiagr} and
\ref{fig:sfevshsls} suggests that the s-process enrichment in this
object may be extrinsic: J053253 is much less luminous than can be
expected from post-AGB evolutionary tracks, and it does not follow the
strong correlation that is found between third dredge-up efficiency
and neutron irradiation. There are, however, two arguments that
contradict this interpretation. First of all, \cite{tomkin89} detected
only two barium dwarfs in their survey of about 200~F-G~dwarfs, which
is comparable to the about one percent of RGB stars that is extrinsic
\citep{macconnell72}. A similar number is hence expected for post-AGB
stars, and it would be very surprising to encounter one in our sample
of only four stars.

The second argument considers the dust that is seen in the system of
J053253. This object clearly has a dust excess, although it is rather
small when compared to the other stars described here. These
excesses are generally not seen in extrinsic objects, as the
circumstellar dust dissipates into the interstellar medium (ISM) on a
timescale that is short compared to the evolutionary timescales. The
relative proximity of the dust in the system of J053253, as follows
from its colour, indicates that, if the material was lost in a wind,
it must have been shed very recently. In this scenario we may
conclude that J053253 is either indeed an intrinsic post-AGB
star, be it at very low luminosity or it is a post-RGB star which
happens to have had a dusty mass-loss event. A post-RGB status
conflicts with internal s-process enhancement.

Another possible explanation for the deviating position of J053253 in
Figs.~\ref{fig:hrdiagr} and \ref{fig:sfevshsls} may be found in the
shape of its SED (see Fig.~\ref{fig:SEDnew}). Although the dust in the
system  is cooler than is typically expected for a
post-AGB object with a circumstellar disc, it radiates the bulk of its
energy at wavelengths shorter than 24~$\mu$m. It is hence a valid
possibility that the circumstellar material surrounding this object
resides in a disc rather than in a freely expanding, detached
shell. It is known from the Galactic sample of post-AGB stars with a
circumstellar disc, that the AGB evolution has been cut short by a
phase of strong binary interaction
\citep{vanwinckel09,gielen11}. Proof of this theory is found in the
orbits that are detected, which are too small to accommodate an AGB
star \citep{vanwinckel09, gorlova12}, and the chemical composition of
the dust, which consists of strongly processed silicates
\citep{gielen11}. This truncation of the stellar evolution should
translate itself in a subluminous object and should also alter the
expected chemical composition of the photosphere: the
[s/$\mathrm{Fe}$] index is supposed to be lower, because fewer third
dredge-ups can take place, while the [hs/ls] index will remain similar
as each third dredge-up brings heavy and light s-process elements to
the surface.

\section{Comparison with AGB nucleosynthetic models}\label{sec:agbmod}

The current theoretical models of the internal nucleosynthesis and the
subsequent photospheric enrichment processes of AGB stars are
sophisticated and rely on the complex interplay between mixing,
nucleosynthesis, and mass loss \citep{busso99,herwig05}. In these models,
extensive nuclear networks have been included \citep[e.g.][]
{cristallo09, cristallo11, karakas10, church09, goriely08, lugaro12},
non-convective mixing like mixing due to differential rotation and thermohaline
mixing have been implemented \citep[e.g.][]{siess03, siess07,
  stancliffe07, angelou11}, the effect of deep mixing or extra mixing
processes has been critically evaluated \citep[e.g.][]{karakas10,
  busso10}, and overshoot regimes have been explored in more detail to
explain the appearance of the $^{13}\mathrm{C}$ pocket
as the main neutron source \citep[e.g.][]{herwig05}. The models are
not self-consistent and overshoot regimes must be incorporated. The
proton ingestion process is complex and detailed simulations at low
metallicities in 3D yield clearly different results as the traditional 1D approaches \citep{stancliffe11}.

Tailored model computations for the individual stars as well as
detailed comparison of the the different codes like we have done in 
\cite{desmedt12},  fall outside the
scope of this paper. The main discrepancy between the model abundances
and the derived ones was that the predicted C/O ratio was much too high compared to the
observed one. Moreover, for that one star in the SMC (J004441.04-732136.4),
 the models fit the s-process {\sl distribution} well, despite
the fact that the absolute abundances were not well matched.
The main conclusion of \cite{desmedt12} was that we
found only a weak dependency of the theoretical predictions on the adopted stellar
evolution code. With the newly analysed objects we want to investigate whether these
discrepancies between AGB nucleosynthesis predictions and the observed
chemical patters are a systematic feature and hence found in more objects.

For this general comparison, we used the nucleosynthetic AGB models from the online-database
FRUITY\footnote{\url{http://www.oa-teramo.inaf.it/fruity}}
\citep[Franec Repository of Upgraded Isotopic Tables \&
Yields,][]{cristallo11}. The models that are currently in the database
focus on low mass AGB stars with initial masses between 1.5 and
3.0~$M_{\odot}$ and metallicities $Z$ in the range from 0.001 to 0.02.

We used the initial mass and metallicity estimates from
Sect.~\ref{ssec:inmass} to determine the model that
approximated these values the best for each object. All models have an
initial mass of 1.5~$M_{\odot}$ as this is the lowest value in the
grid. As we suspect that J053253 is extrinsically enriched, we
assume that the donor has this mass as well (see Sect.~\ref{ssec:inmass}). We
use a metallicity of $Z = 0.001$ for J050632, J052043, and J053250 as
also this value is on the lower edge of the grid, and $Z = 0.003$ and
0.006 for J053253, because the metallicity of this object falls
between these two values. 

\begin{figure}
\resizebox{\hsize}{!}{
\includegraphics{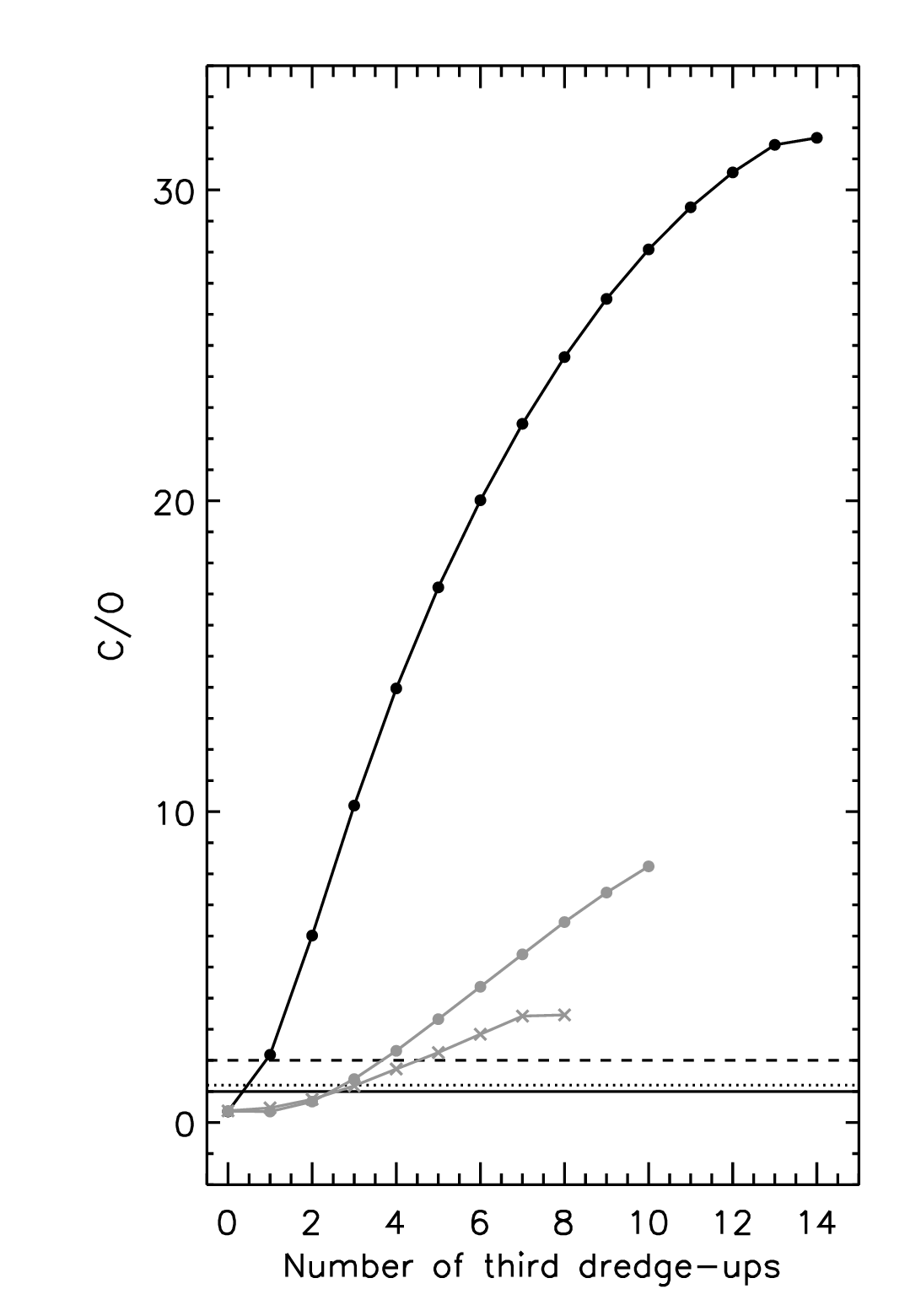}
\includegraphics{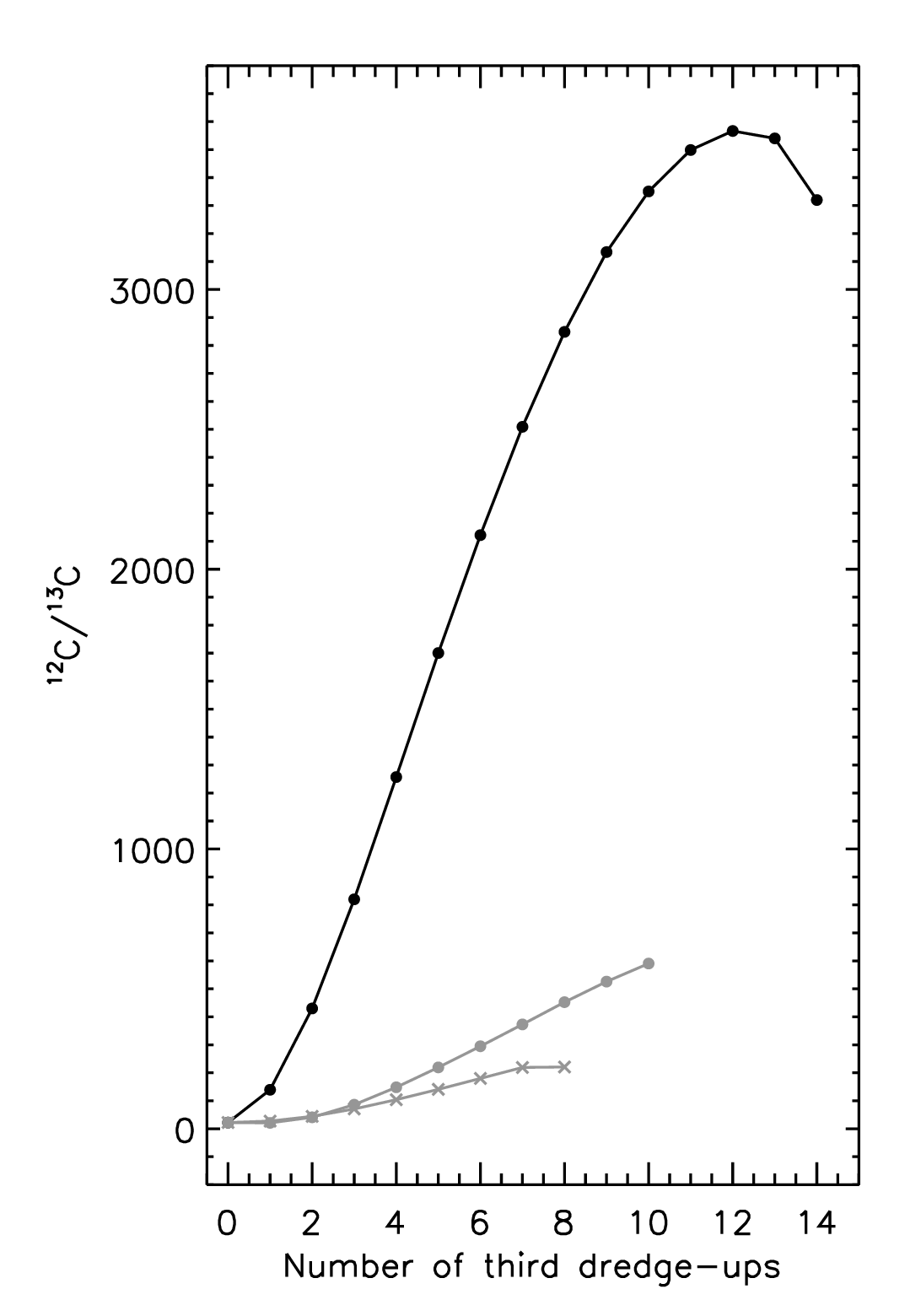}
}
\resizebox{\hsize}{!}{
\includegraphics{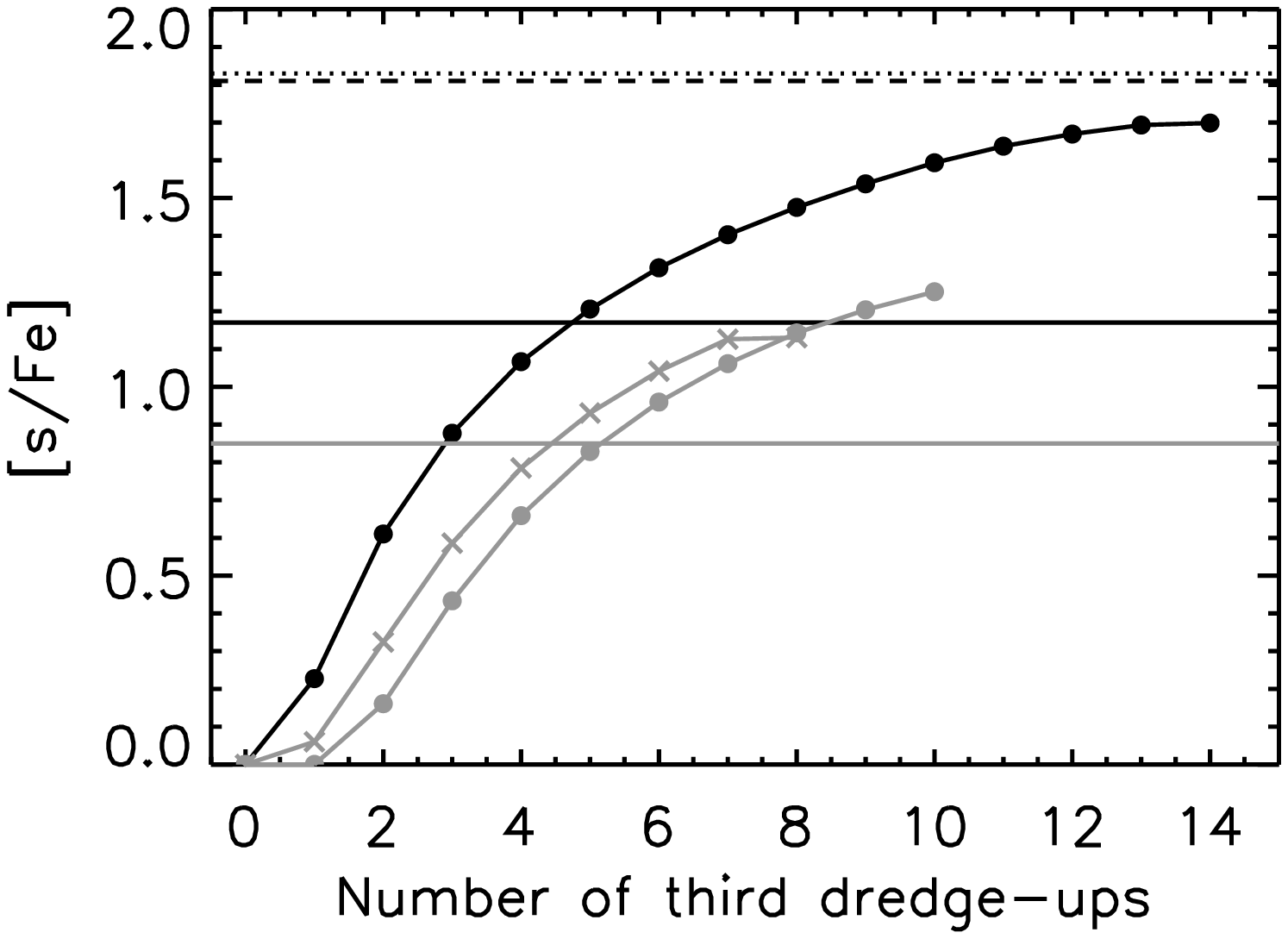}
\includegraphics{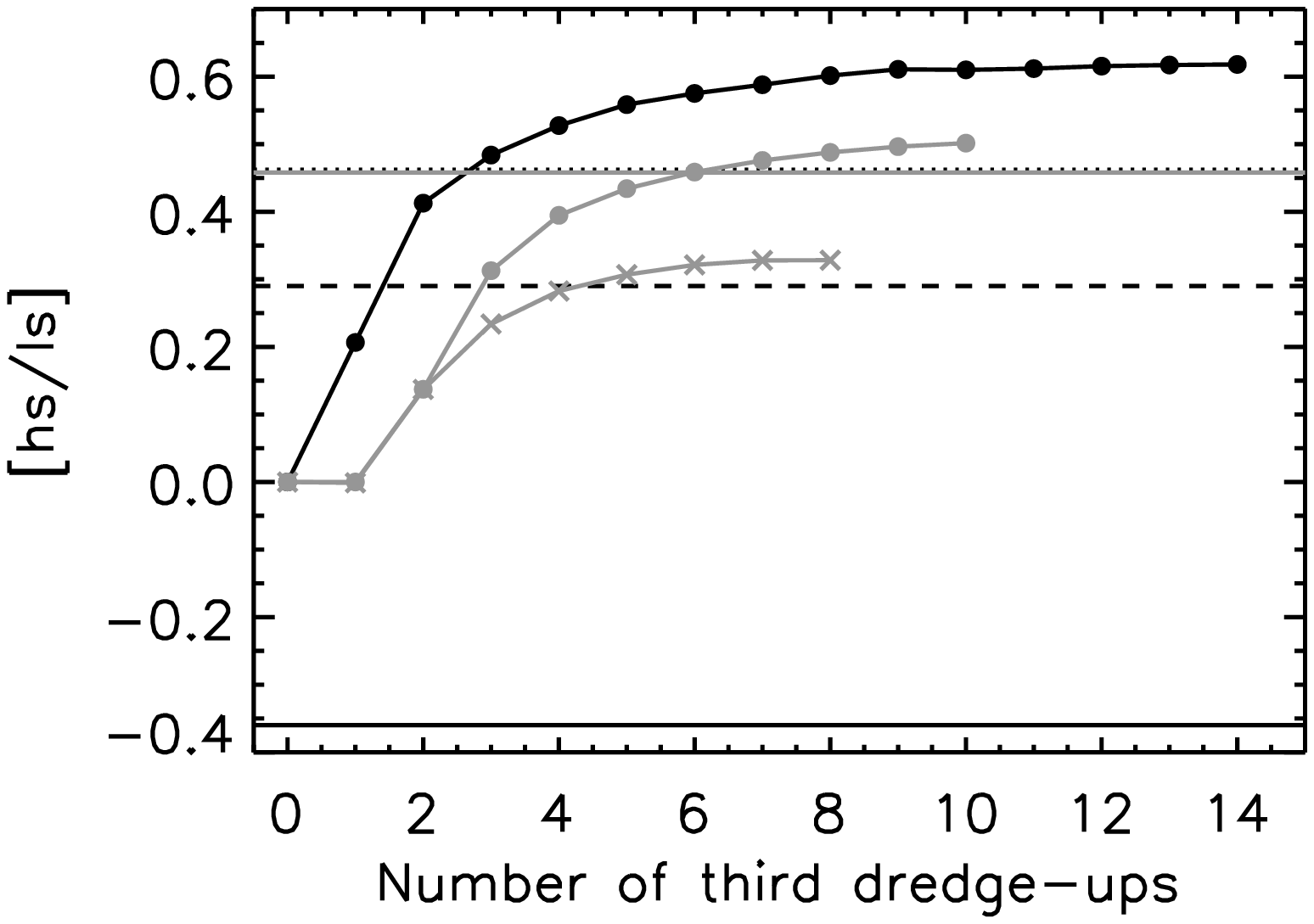}
}
\caption{
Some characteristics of the used nucleosynthetic AGB models. We show the $\mathrm{C}$/$\mathrm{O}$ and $^{12}\mathrm{C}$/$^{13}\mathrm{C}$ ratio, and the [s/$\mathrm{Fe}$] and [hs/ls] indexes as a function of the number of third dredge-ups the model has undergone. Each third dredge-up episode is represented by a dot or a cross, depending on the nucleosynthetic model, and they are linked for clarity. The model with metallicity $Z = 0.001$ is shown as connected black dots, while the models with metallicity $Z = 0.003$ and $Z = 0.006$ are represented by grey dots and crosses respectively. 
The values obtained for the stars in our sample are overplotted as
horizontal lines in the same colour (black,grey) as their most appropriate nucleosynthetic model. J050632 is represented by a full, black line, J052043 by a dashed, black line, J053250 by a dotted, black line, and J053253 by a full, grey line.}
\label{fig:AGBmod_stat}
\end{figure}

In Fig.~\ref{fig:AGBmod_stat} we show some characteristics of the
models as a function of the number of third dredge-up episodes
the star has undergone. The model with metallicity $Z = 0.001$ reaches
the post-AGB phase after fourteen third dredge-ups, while this already
happens after respectively ten and eight for the models with $Z =
0.003$ and 0.006. We plotted the $\mathrm{C}$/$\mathrm{O}$ and
$^{12}\mathrm{C}$/$^{13}\mathrm{C}$ ratio, as well as the
[s/$\mathrm{Fe}$] and [hs/ls] indexes as a function of the number of
dredge-ups the star has undergone at that point.

Also for these stars, we confirm that the relative abundances 
of both $\mathrm{C}$ and $\mathrm{O}$ are
badly predicted by the models:
$\mathrm{C}$ is in general overpredicted while the model values
for $\mathrm{O}$ remain well below the actual determined value. This
is a fundamental problem of the models, as especially for the
$\mathrm{C}$-abundances, the model values lie well outside the error
bars on our relative abundances.
The combination of both deviations leads to a too high predicted
$\mathrm{C}/\mathrm{O}$ ratio of the AGB models. From the first panel
in Fig.~\ref{fig:AGBmod_stat}, it follows that the model with $Z =
0.001$ already estimates the corresponding observed
$\mathrm{C}/\mathrm{O}$ ratios after the occurrence of only one third
dredge-up. The $\mathrm{C}/\mathrm{O}$ ratio of the final model with
this metallicity overrates the observed one by a factor $\sim$20. 
As in \cite{desmedt12}, the high observed
$\mathrm{O}$-abundance, which also contributes to the
$\mathrm{C}/\mathrm{O}$ ratio, is not accounted for.

Our spectra do not allow deduction of the
$^{12}\mathrm{C}$/$^{13}\mathrm{C}$ isotopic fractions, because the
$\mathrm{CN}$ violet and red systems around 3875 and 6191~\AA{} are
not detected at the S/N of our spectra and the temperature of our
stars. Our spectral range does not cover the additional
$\mathrm{CN}$ red system at 6945~\AA{}, and the wavelength region from
1-5~$\mu$m, where several $\mathrm{CO}$ forests could be
found.  Determining the $^{12}\mathrm{C}$/$^{13}\mathrm{C}$ isotopic
ratio is an observational challenge in these objects but it would
allow a very good test on the predictive power of the AGB models.




For the objects with the highest s-process enhancement, J052043 and
J053250, the comparison of the s-process distribution with theoretical
models matches relatively well, but the nucleosynthetic models underestimate
the actual abundances of the light s-process elements yttrium
($\mathrm{Y}$, $Z = 39$) and zirconium ($\mathrm{Zr}$, $Z = 40$) by
about 0.2~dex. The underestimation of the observed relative
abundances of the light s-process elements leads to an overestimation
of the actual [hs/ls] index, which forms an indication of the neutron
irradiation. In the fourth panel of Fig.~\ref{fig:AGBmod_stat}, it can
be seen that the models already overestimate this index after the
second occurrence of a third dredge-up for J052043, and after the
third for J053250. The overall enhancement in s-process elements, or
the efficiency of the third dredge-up, on the other hand, is
relatively well reproduced for both objects (see the third panel of
Fig.~\ref{fig:AGBmod_stat}) after the full series of predicted 3rd
dredge-up events.

For the other two objects, J050632 and J053253, the comparison with
the theoretical models is not as good. 
For J053253 the observed abundances are systematically too low with
respect to the standard model (Fig.~\ref{fig:AGBmod_stat}). This is another indication that
the s-process and the third dredge-ups are indeed inhibited in some
way (see Sect.~\ref{sec:J053253extr}).

The comparison for J050632 points to a systematic difference: the
model underestimates the light s-process abundances while it heavily
overestimates the observed abundance pattern for the heavy s-process
elements. This results in  a negative [hs/ls]
index, which is not predicted by the models (see
Fig.~\ref{fig:AGBmod_stat}).

Although J050632, J052043 and J053250 are intrinsically enriched
post-AGB objects with a very similar metallicity, luminosity, and
hence initial mass, they display quite different levels of overabundance as
well as a significantly different s-process distribution. It therefore
follows that the AGB models have to be individually adapted to match
the observed abundance patterns.

%

\section{Conclusions}\label{sec:concl}

In \cite{vanaarle11}, we classified the objects J050632,
J052043, J053250, and J053253 (see Table~\ref{table:samplestars}) as
spectroscopically confirmed post-AGB stars, because of their spectral
type, IR colours, and estimated luminosity. The chemical analysis we
carried out in this paper was based on high-resolution UVES spectra,
and confirmed their post-AGB status, as the third dredge-up clearly
took place in all objects: they are $\mathrm{C}$-rich, and enhanced in
s-process elements. We deduced abundances of heavy s-process elements
for all stars in the sample, and quantified also abundances of
  elements well beyond the Ba-peak. The low S/N of the spectra in the
blue spectral domain, where the lines of heavy s-process elements
abondan, prevented very accurate determinations.

The metallicity of all stars except J053253 is
considerably lower than the average value that is observed for the
LMC. This is in accordance with the possibility that these objects are
old, which means that they should have a low initial mass. The latter
is also confirmed by their low luminosities. We estimate the initial
mass of the three objects to be between 1 and 1.5~M$_{\odot}$.

All sample stars except J053253 confirm the correlation
between the efficiency of the third-dredge up and the neutron exposure
that is present in the Galactic sample of post-AGB objects. The
non-existence of a correlation between metallicity and neutron
irradiation is also further corroborated.

The subluminosity of J053253, its small but warm infrared
excess, and the deviation of the s-process abundance pattern from the
expected trend (see Fig.~\ref{fig:sfevshsls}) are all observational
indications that this object is not evolving on a single-star
evolutionary track. We argued that the dust in the system of this star
may reside in a disc rather than in an freely expanding, detached
shell. This would mean that this object is likely a binary. From
Galactic binary post-AGB stars with a disc, it is known that the AGB
evolutionary phase in these objects is truncated, because of the
detected orbits \citep{vanwinckel09}. This can explain the subluminosity and low
[s/$\mathrm{Fe}$] index that are observed for this star, but radial
velocity time-series are badly needed.

The 21~$\mu$m feature has been detected in the spectrum of
J052043 \citep{volk11}. As the fifteen Galactic objects
that display this feature are post-carbon stars which are strongly
enhanced in s-process elements \citep{hrivnak08}, this detection shows
that this is also valid for the LMC. Similar LMC 21~$\mu$m sources are
known \citep{volk11}, and it would be useful to obtain abundances for
them as well. In the SMC, there is only one known 21~$\mu$m source:
J004441.04-732136.4 \citep{volk11}, which was recently subjected to a
detailed chemical study \citep{desmedt12}. This object has an initial
mass of$\sim$1.3~M$_{\odot}$, a metallicity of -1.3, and turned out to be
one of the most s-process enriched objects known to date. A more
detailed comparison with model predictions using two different independent codes
shows that very similar conflicts exist between the deduced and
predicted abundances as found in this contribution.

From the comparison with AGB models, we
conclude that the available AGB models overestimate the observed
$\mathrm{C}$/$\mathrm{O}$ ratios and fail to reproduce the variety of
[hs/ls] indexes that is observed in otherwise very similar
stars. This seems to be a common feature \citep{desmedt12}.
Additional near infrared spectra would be useful to determine
the $^{12}\mathrm{C}$/$^{13}\mathrm{C}$ ratio of our stars, as this
ratio is predicted to be very dependent on the number of dredge-up
events the star has suffered on the AGB. The $\mathrm{Pb}$ production
is predicted to be very high in these metal poor objects but an
observational constraint will require additional high S/N blue spectra.

This paper is the first systematic study of enriched post-AGB stars in
the LMC. Because the range of luminosities of this subsample is rather
small, more general and systematic confrontations of the observed
abundances with AGB nucleosynthetic models will only become possible
when the abundances and characteristic parameters of other objects
covering a wider range of metallicities and luminosities have been studied. We
therefore initiated a large systematic low-resolution survey \citep{kamath13}
to identify the best suitable candidates both in the LMC and SMC which
will enable us to quantify a wide range of chemical species in
post-AGB stars covering a wide luminosity range.

Despite the
small range in luminosities, the detected abundances cover a range in
s-process overabundances and abundance patterns which call for very
detailed, tailored models that will match the derived abundance
patterns individually. Only such a systematic study can make significant
progress in the understanding of the complex interplay between mixing,
nucleosynthesis and mass-loss which characterise the final evolution
of solar-like stars, and is the ultimate goal of our research.

\begin{acknowledgements}
E. van Aarle acknowledges support from the Fund for Scientific Research of
Flanders (FWO) under  grant number G.0470.07. 
The authors thank inspiring discussions with Tom Lloyd Evans, Lionel
Siess, Stephane Goriely, Amanda Karakas, and Nadya Gorlova.

\end{acknowledgements}


\end{document}